\documentclass[usenatbib]{mnras}
\usepackage{mathptmx}
\usepackage{color}
\usepackage{xcolor}
\usepackage{graphics,graphicx,amssymb,amsmath}
\usepackage{multirow}
\usepackage{soul}
\soulregister\citep7
\soulregister\citet7
\soulregister\ref7

\def\gtrsim{\lower 2pt \hbox{$\, \buildrel {\scriptstyle >}\over
{\scriptstyle \sim}\,$}}
\def\lesssim{\lower 2pt \hbox{$\, \buildrel {\scriptstyle <}\over
{\scriptstyle \sim}\,$}}
\def\Spitzer{{\sl Spitzer}}

\def\hst{{\sl HST}}

\def\xmm{{\sl XMM-Newton}}

\def\chandra{{\sl Chandra}}

\def\hst{{\sl HST}}

\def\xsb{PJ0116-24}
\def\xsa{PJ1336+49}
\def\xsc{PJ1053+60}

\def\approxlt{\lower.2em\hbox{$\buildrel < \over \sim$}}
\def\approxgt{\lower.2em\hbox{$\buildrel > \over \sim$}}

\def\sous{HyLIRGs}
\def\sou{HyLIRG}
\def\xp{$X_{\sc HMXB}$}
\def\ins{{\sl XMM-Newton}}

\def\planck{{\sl Planck}}

\date{Accepted XXX. Received YYY; in original form ZZZ}

\pubyear{2020}

\begin{document}
\title[]{X-ray detection of the most extreme star-forming galaxies at the cosmic noon via strong lensing}
\author[Q. D. Wang et al.]{Q. Daniel Wang$^{1}$\thanks{Contact e-mail:wqd@umass.edu}, Carlos Garcia Diaz$^{1}$, Patrick S. Kamieneski$^{2}$, Kevin C. Harrington$^{3}$, Min S. Yun$^{1}$,
\newauthor{Nicholas Foo$^{2}$, Brenda L. Frye$^{4}$, Eric F. Jimenez-Andrade$^{5}$, Daizhong Liu$^{6}$, James D. Lowenthal$^{7}$,}
\newauthor{Bel\'{e}n Alcalde Pampliega$^{3}$, Massimo Pascale$^{8}$, Amit Vishwas$^{9}$, \& Mark A. Gurwell$^{10}$}\\
$^{1}$ Department of Astronomy, University of Massachusetts, Amherst, MA 01003, USA\\
$^{2}$ School of Earth and Space Exploration, Arizona State University, Tempe, AZ 85287-6004, USA\\ 
$^{3}$ European Southern Observatory, Alonso de C\'{o}rdova 3107, Viticura, Casilla 19001, Santiago de Chile, Chile\\
$^{4}$ Department of Astronomy/Steward Observatory, 933 North Cherry Avenue, University of Arizona, Tucson, AZ 85721, USA\\
$^{5}$ Instituto de Radioastronomía y Astrofísica, Universidad Nacional Aut\'{o}noma de M\'{e}xico, Antigua Carretera a P\'{a}tzcuaro \# 8701,\\ Ex-Hda. San Jos\'{e} de la Huerta, Morelia, Michoacán, M\'{e}xico C.P. 58089\\
$^{6}$ Max-Planck-Institut f\"{u}r extraterrestrische Physik, Giessenbachstrasse 1
Garching, Bayern, DE 85748, Germany\\
$^{7}$ Smith College, Northampton, MA 01063, USA\\
$^{8}$Department of Astronomy, University of California, Berkeley, CA 94720, USA\\
$^{9}$ Department of Astronomy, Cornell University, Space Sciences Building, Ithaca, NY 14853, USA\\
$^{10}$ Center for Astrophysics $\vert$ Harvard \& Smithsonian, 60 Garden Street, Cambridge, MA 02138, USA\\
 }
\maketitle
\begin{abstract}
Hyper-luminous infrared galaxies (HyLIRGs) are the most extreme star-forming systems observed in the early Universe, and their properties still elude comprehensive understanding. We have undertaken a large XMM-Newton observing program to probe the total accreting black hole population in three HyLIRGs at z = 2.12, 3.25, and 3.55, gravitationally lensed by foreground galaxies. Selected from the {\sl Planck} All-Sky Survey to Analyze Gravitationally-lensed Extreme Starbursts (PASSAGES), these HyLIRGs have apparent infrared luminosities  $> 10^{14} {\rm~L_{\odot}}$. Our observations revealed X-ray emission in each of them. \xsa\ appears to be dominated by high-mass X-ray binaries (HMXBs). Remarkably, the luminosity of this non-AGN X-ray emission exceeds by a factor of about three the value obtained by calibration with local galaxies with much lower star formation rates. This enhanced X-ray emission most likely highlights the efficacy of dynamical HMXB production within compact clusters, which is an important mode of star formation in HyLIRGs. The remaining two (\xsb\ and \xsc) morphologically and spectrally exhibit a compact X-ray component in addition to the extended non-AGN X-ray emission, indicating the presence of Active Galactic Nuclei (AGNs). The AGN appears to be centrally located in the reconstructed source plane images of \xsb, which manifests its star-forming activity predominantly within an extended galactic disk. In contrast, the AGN in the field of \xsc\ is projected 60 kpc away from the extreme star-forming galaxy and could be ejected from it. These results underline the synergistic potential of deep X-ray observations with strong lensing for the study of high-energy astrophysical phenomena in HyLIRGs.

\end{abstract}
\begin{keywords}
galaxies: high-redshift, galaxies: nuclei, galaxies: starburst, X-rays: galaxies, X-rays: binaries, gravitational lensing: strong
\end{keywords}

\section{INTRODUCTION}
\label{s:intro}

Hyperluminous infrared galaxies (\sous) are the brightest galaxies in the Universe and represent the most extreme version of dusty star-forming galaxies (DSFGs) \citep[e.g.][]{Irwin1998,Rowan-Robinson2000,Blain2002,Casey2014}. Identified by their large intrinsic infrared (IR) luminosity $> 10^{13} {\rm~L_\odot}$, \sous\ are typically found at $z \gtrsim 2$ and are rare (a few per deg$^{2}$ according to \cite{Gruppioni2013}). If the IR luminosity is due to star formation (SF), then the corresponding SF rate (SFR) of such a \sou\ is $\gtrsim 10^{3} {\rm~M_\odot~yr^{-1}}$. However, the IR luminosity could be significantly contributed by an Active Galactic Nucleus (AGN), as observed especially at $z < 1$ \citep[e.g.,][]{Farrah2002,Ruiz2007}. In this case, the galaxy may be detected as a so-called Hot Dust-Obscured Galaxy \citep[Hot DOG; e.g.,][]{Wu2018}. In any case, \sous\ are natural laboratories for studying the physical processes that govern the rapid growth and quenching of massive young galaxies and/or the co-evolution between SF and supermassive black hole (SMBH) accretion in the early Universe \citep[e.g.][]{Blain2002,Casey2014}. Since there are no local analogs to their large inferred molecular gas mass (up to $10^{11} M_\odot$; \cite{Harrington2018,Harrington2021}) and gas mass fraction (40-80\%), as well as their extreme luminosity \citep[e.g.,][]{Carilli2013}, a detailed multi-wavelength study of \sous\ is the only way to probe the underlying physical processes.  

\subsection{Studying \sous\ in the high-energy astrophysics context}\label{ss:intro-HEA}
High-energy astrophysical processes are thought to play a central role in regulating the formation and evolution of galaxies and even their large-scale environment. Such processes are mostly associated with SF and SMBH accretion, which are probably coordinated processes, especially during the early rapid co-evolution stage \citep[e.g.,][]{Hopkins2008,Mancuso2016}. However, due to the limited sensitivity and resolving power of existing X-ray observatories -- a primary tool for studying high-energy phenomena, it remains unclear how this regulation works and whether relevant key scaling relations derived locally can be applied, for example, to high-$z$ extreme star-forming galaxies evolving under very different conditions. 
Here we focus on two key questions: 1) How might the ratio of the non-AGN X-ray luminosity (dominated by high-mass X-ray binaries or HMXBs) to SFR, $L_X/{\rm SFR}$ (hereafter \xp), in \sous\ differ from that in more normal star-forming galaxies? 2) How do AGNs co-evolve with SF in HyLIRGs? 

\subsubsection{Uncertainty in the \xp\ calibration}\label{sss:intro-ratio}

The calibration of the \xp\ factor is important for multiple reasons. First, if \xp\ is known, then one may use it to infer the SFR of a galaxy from its $L_X$, which may be measured more reliably than other SF tracers, especially when dust obscuration is severe \citep[e.g.,][]{Ghosh2001}. However, when the AGN contribution is potentially significant, $L_X$ gives only an upper limit to the SFR \citep[e.g.,][]{Mineo2012,Mineo2014}. Second, one may infer the presence of an AGN in a galaxy if its $L_X$ is significantly greater than that estimated from the SFR (if independently measured), together with a known \xp. Third and probably more importantly, the determination of a systematic \xp\ variation with certain galaxy properties could provide deep insights into the formation process of non-AGN X-ray sources. In active star-forming galaxies, for example, $L_X$ is mainly determined by HMXBs, especially so-called ultra-luminous X-ray sources (ULXs) with $L_{\rm 0.5-10~keV}. \gtrsim 1 \times 10^{39} {\rm~ergs~s^{-1}}$ \citep[e.g.,][]{Kaaret2014,Lehmer2021}. Here we consider HMXBs to include intermediate-mass X-ray binaries with donor stars whose lifetimes are up to a few $\times 10^8$~yrs \citep[e.g.,][]{Poutanen2013,Hunt2023} -- the characteristic starburst age of \sous. Each HMXB consists of a compact object (neutron star or black hole) and a donor star whose current mass is greater than $3 {\rm~M_\odot}$\footnote{while one with a donor star whose current mass is less than 3${\rm~M_\odot}$ is called a low-mass X-ray binary or LMXB \citep[e.g.,][]{Fragos2013}}. The black hole (BH) may have stellar mass ($< 10^2 {\rm~M_\odot}$) or intermediate mass (IMBH with the BH mass $\gtrsim 10^2 {\rm~M_\odot}$ but $< 10^5 {\rm~M_\odot}$). 

Extensive calibrations of \xp\ have been performed primarily with X-ray observations of nearby star-forming galaxies \citep[e.g.,][]{Fabbiano1985,Bauer2002,Grimm2003,Ranalli2003,Persic2007,Mineo2012,Riccio2023}. Both spatially resolved and integrated measurements of $L_X$ are used. 
Recent calibrations have typically been performed in the 0.5-8~keV band, including the conversion of luminosities estimated in other bands, assuming a spectral model (e.g., photon index $\Gamma \approx 2$ and $N_{H} \approx 3 \times 10^{21} {\rm~cm^{-2}}$; \cite{Mineo2012,Mineo2014,Lehmer2021}). 

\begin{figure}
\centerline{
\includegraphics[width=1.0\linewidth,angle=0]{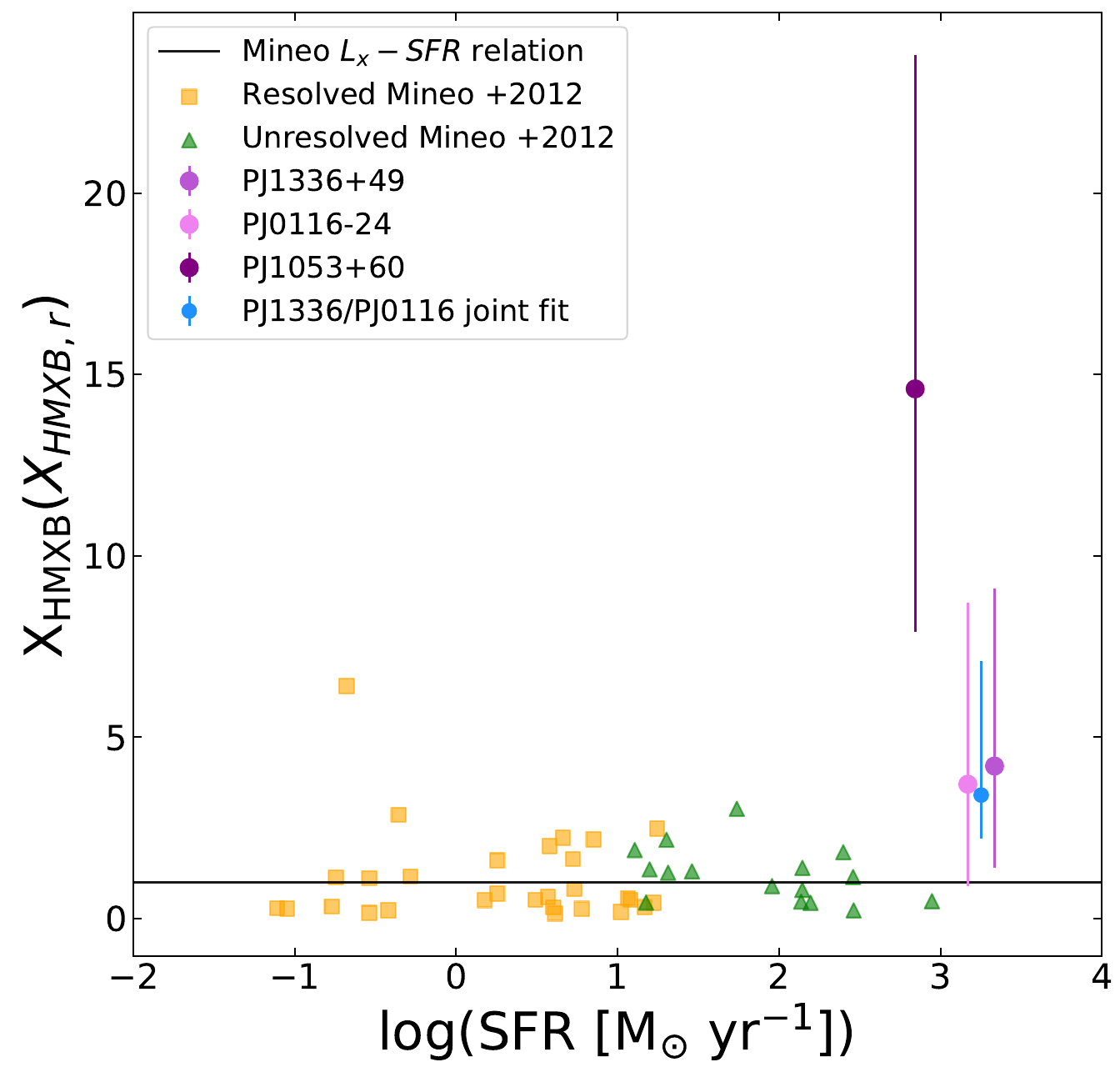}
}\caption{\xp\ data for individual galaxies included in the study by \citet{Mineo2012}, normalized to their best-fit mean or to the coefficient of the $L_X-$SFR relation (as given in Eq.~\ref{e:xsfr}) for those galaxies with resolved X-ray emission. Our \xp\ measurements for the three \sous\  (Table~\ref{t:spec}) with the SFR corrected for the lensing magnifications are included for comparison.
}
\label{f:f1}
\end{figure}

The probably most commonly used calibration is obtained by \citet{Mineo2012} primarily through a careful spatially resolved X-ray study of star-forming galaxies within a distance of $< 40$~Mpc. In these galaxies, the HMXB contributions are statistically determined.  The  best-fit $L_X-$SFR relation \citep{Mineo2012} is
\begin{equation} 
L_X ({\rm erg~s^{-1}})= 2.61 \times 10^{39}  {\rm~SFR(M_{\odot}~yr^{-1})},
\label{e:xsfr} 
\end{equation} 
which is equivalent to a \xp\ calibration (Fig.~\ref{f:f1}).
 The scatter around this relation (or \xp\ calibration) is large (RMS $= 0.43$ dex), at least partly due to both the limited sampling of the star formation history and the variation in the number and luminosity of the brightest sources \citep{Mineo2012}. The SFRs of the local sample galaxies are all $< 20 {\rm~M_\odot~yr^{-1}}$ and have a median value of only $5.3 {\rm~M_{\odot}~yr^{-1}}$, leading to large stochastic uncertainties. 
{\sl For comparison}, as in \citet{Mineo2012}, Fig.~\ref{f:f1} also includes the \xp\ data obtained for more distant galaxies for which only the integrated total X-ray luminosities are measured, extending the SFR range to higher values by about two orders of magnitude. 

The inclusion of such X-ray-unresolved integrated measurements in the calibration of \xp\  would introduce substantial systematic effects  \citep[e.g.,][]{Mineo2014,Riccio2023}. In fact, using only the unresolved sample galaxies yields a coefficient value of $\sim 1.5$ higher than in Eq.~\ref{e:xsfr}. This statistical X-ray emission excess is also apparent for the data points with SFRs $> 20 {\rm~M_\odot~yr^{-1}}$ in Fig.~\ref{f:f1} and is expected because of the inclusion of unresolved X-ray contributions (e.g., from diffuse hot gas, LMXBs, and AGNs;  \citet{Mineo2012}). 
 The contamination from diffuse hot plasma is typically important at $\lesssim 1.5$~keV \citep[e.g.,][]{Lehmer2022,Wang2021} and is thus sensitive to both the X-ray absorption and the actual energy band used to measure the X-ray emission. AGNs are present in $\sim 25-50\%$ of galaxies with high SFRs \citep{Kirkpatrick2017} and contribute more at higher energies. Such varying contamination remains difficult to quantify, but could contribute significantly to the RMS of the galaxies, as well as the bias toward an over-estimated \xp\ in such a calibration.

The use of a different IMF must also be considered. We here use the IR luminosity to SFR conversion that assumes the Kroupa IMF \citep{Kennicutt1998,Calzetti2013}. This IMF is more realistic than the Salpeter IMF assumed to get Eq.~\ref{e:xsfr} when low-mass stars (down to $0.1 {\rm~M_\odot}$) are also considered, which contribute the bulk of the stellar mass. To account for this difference, we apply a correction factor of 1.5 \citep{Madau2014} to the aforementioned calibration, resulting in the following Kroupa IMF-based calibration,
\begin{equation} 
X_{\sc HMXB,r}   \approx 3.9 \times 10^{39} {\rm~erg~s^{-1}/(M_{\odot}~yr^{-1}}), 
\label{e:xsfr-r} 
\end{equation} 
which we use as the reference or unit for our measurement of \xp. The use of this calibration, instead of the one based on the total X-ray luminosity of a galaxy in the 0.5-8~keV range \citep{Mineo2014}, is appropriate here because we measure \xp\ or the apparent non-AGN luminosity $\mu L_X$ by fitting the spectra of our sample \sous\ with $z > 2$ in the rest-frame energy range above 1.5~keV (\S~\ref{s:res}), where the diffuse hot plasma contamination can be neglected. The LMXB contribution should also be negligible in such extreme star-forming galaxies. 

There are other systematics that affect existing \xp\ calibrations, most of which are based on the total X-ray luminosities of galaxies.
Both theoretical argument and observational evidence suggest that \xp\ increases with decreasing metallicity and/or decreasing total age of stellar populations \citep[e.g., ][]{Fragos2013,Kaaret2014,Lehmer2021}. In addition, the dynamical formation of HMXBs in young ($\lesssim 10$~Myr) or intermediate-age (10-400 Myr) dense stellar clusters \citep[e.g.,][]{Garofali2012,Kremer2020,Rizzuto2022,Hunt2023} could also play an important role, especially in \sous\ (see \S~\ref{ss:dis-Xf}). 
Furthermore, a general deficiency in X-ray emission from (ultra)luminous infrared galaxies, or (U)LIRGs, relative to their $L_{FIR}$ or SFR has long been noted \citep[e.g.,][]{Persic2007,Iwasawa2009,Lehmer2010,Torres-Alba2018}. The most extreme sources are 1-2 dex below the $L_{\rm 2-10~keV}-$ SFR relation \citep{Ranalli2003}. 
This faintness is likely caused by substantial intrinsic X-ray absorption and/or the extreme youth of compact starbursts in such local extreme galaxies. The required $N_H$ would be on the order of $\gtrsim 10^{24} {\rm~cm^{-2}}$. The compactness of starbursts in (U)LIRGs is also related to their extreme youth (with starburst ages less than a few Myr; \citet{Inami2013, Diaz-Santos2017,West2023}), in which massive stars contribute strongly to the IR emission, but the number of HMXBs is small. Alternatively, the compactness could lead to the rapid destruction of clusters \citep[e.g.,][]{Linden2021,Linden2023}. These effects are still difficult to quantify for (U)LIRGs. Furthermore, recent X-ray stacking analyses of ``normal'' galaxies up to $z \sim 2.4$ (e.g. using data from the Chandra Deep Field and Chandra COSMOS surveys) suggest that \xp\ increases moderately (e.g, by a factor of $\sim 2$) with increasing redshift and decreasing metallicity, although the confusion with low luminosity AGNs can be serious \citep[e.g.,][]{Kaaret2014,Lehmer2019,Fornasini2019,Fornasini2020,Lehmer2021,Lehmer2022,Gilbertson2022}. 

These statistical and systematic effects on the existing calibrations illustrate the importance of a {\sl spatially resolved} measurement of \xp, preferably in a harder energy band and with a sufficiently large total SFR. Incidentally, this measurement could also help us to determine how BH binaries of masses $\gtrsim 30-60 {\rm~M_{\odot}}$ may be formed, which have been detected via their gravitational wave emission from coalescence \citep[e.g.,][]{Abbott2020,Belczynski2020}. 

\subsubsection{Uncertainty in the AGN and \sou\ co-evolution}\label{sss:intro-AGN}
To date, information on the presence of AGNs in \sous\ at $z \gtrsim 2$ is very limited. While some \sous\ have been identified as Hot DOGs via their enhanced mid-IR emission observed in Wide-field Infrared Survey Explorer ({\sl WISE}) bands \citep[e.g.,][]{Wu2018}, they probably represent only a special AGN accretion stage and do not tell a complete story about the SF and SMBH co-evolution in such galaxies.  An intriguing example is the \xmm\ detection of the unlensed \sou, HATLAS J084933.4+021443 at $z=2.41$  \citep{Ivison2019}. Its \xmm\ spectrum suggests the presence of an AGN, with its high X-ray luminosity indicating a potential energy release that could rival the starburst. Interestingly, however, the apparent X-ray absorption ($N_H \lesssim 5 \times 10^{21} {\rm~cm^{-2}}$) suggests that this AGN is not deeply embedded. Because the \xmm\ data provide little spatial information about the X-ray emission from this unlensed \sou, its actual physical relationship to the AGN is not clear. Several scenarios remain to be tested. One possibility is that the AGN is observed through a cavity of the galaxy it excavates. Alternatively, if the AGN is not co-spatial with the starburst, it may have been ejected from the galaxy or has evolved into the naked quasar phase in a separate faint companion galaxy.

\begin{table*}
\begin{center}
\caption{Parameters of the targets}
\label{t:targets}
{
\begin{tabular}{lcccccccccr}
\hline\hline
Target & R.A. (J2000) & Dec. (J2000) & $z_{ DSFG}$ &$\mu$SFR&{\sl WISE} W1 & $z_{lens}$ & $\mu$ & $\theta_E$ & $M_{<\theta_E}$ & $N_{G,H}$$^d$  \\ 
Name & (h~~m~~s) & ($^\circ~~~^{\prime}~~~^{\prime\prime}$) &       &${\rm 10^{4} 
M_\odot~yr^{-1}}$& mag &   & & $^{\prime\prime}$ (kpc)&${\rm~10^{11}~M_\odot}$ &$10^{20} {\rm~cm^{-2}}$ \\ 
\hline
\xsa & 13 36 34.9 & +49 13 14&3.2541 &1.8 & 14.8 &0.26	& 8.3&	1.2 (4.9)& $1.7 \pm 0.2$& 0.83 \\ 
\xsb &01 16 46.8 & -24 37 02 &2.1249 &2.5$^a$ & 13.8&0.555$^b$ 
&17$^a$ &2.4 (16) & $15.4 \pm 0.3$&1.6 \\ 
\xsc &10 53 22.6 & +60 51 47 &3.5490 &1.7 & 15.5&0.837 &24$^c$ & 5.9 (46) & 77$^c$ & 0.96 \\ 
\hline		
\end{tabular} 
}
\end{center}
{
Listed target parameters are the \planck\ source name, centroid position, redshift ($z_{ DSFG}$), 
apparent SFR ($\mu$SFR), and {\sl WISE} band 1 magnitude of each \sou; the redshift of the foreground lensing galaxy ($z_{lens}$), mean lensing magnification ($\mu$), Einstein radius ($\theta_E$), enclosed gravitational mass ($M_{<\theta_E}$); and Galactic HI column density ($N_{G,H}$). Most of the parameter values are obtained from \cite{Berman2022,Kamieneski2023}, except for $^a$ from an updated spectral energy distribution (SED) fit (including data from ALMA observations) \citep{Liu2023}, $^b$ from an observation made with the Very Large Telescope (VLT)/MUSE \citep{Liu2023} and 
$^c$ from Appendix~\ref{a:PJ1053-lens-model}, and $^d$ as detected by the 21-cm survey \citep{HI4PI} and obtained from the website \url{https://heasarc.gsfc.nasa.gov/cgi-bin/Tools/w3nh/w3nh.pl} at the positions of each source. 
}
\end{table*}

\subsection{X-ray \sous\ through strong gravitational lensing foregrounds}\label{ss:intro-lensing}

We have been exploring the use of strong gravitational lensing - nature's magnifying glass - to overcome the observational limitations in observing dusty star-forming galaxies (DSFGs), especially \sous\ \citep[e.g.,][]{Harrington2021,Berman2022,Kamieneski2023}. 
Strong lensing provides a unique opportunity to amplify the flux of a galaxy, making it easier to detect. In addition, the increased angular size potentially allows us to resolve its structure and separate different components, while an AGN that remains point-like in multiple images can then be relatively easily identified. Existing studies have already shown that \sous\ can be complex systems of multiple components: e.g., a dominant DSFG accompanied by separate galaxies, which may or may not be physically associated.
Aided by the strong lensing, we can have a better chance to pinpoint the responsible component for an observed signature. 

We report here initial results from a large \ins\ program observing three strongly lensed \sous\ at the redshifts 2.12, 3.25, and 3.55 (Table~\ref{t:targets}) with a total exposure of 526~ks (Table~\ref{t:obs}). The excellent sensitivity and broad energy coverage of \ins, combined with the intrinsic brightness and lensing magnification of \sous, provide us with a unique opportunity to probe the high-energy activity in these extraordinary galaxies. The observations are sensitive to HMXBs and AGNs in the $\sim 1.5-36$ keV rest frame. In this energy range, both the diffuse emission contamination and the X-ray absorption of the interstellar medium (ISM) are substantially smaller than in the 0.5-1.5~keV band \citep[e.g.,][]{Lehmer2022,Wang2021}. We may reasonably assume that the HMXB emission statistically follows the rest-frame far-IR emission since both typically trace recent massive SF on similar spatial and temporal scales (down to $\sim 1$~kpc and $10^7$~yr; \citet{Persic2007,Mineo2012,Cochrane2019}). Under this assumption, both the observed X-ray and dust continuum emissions are expected to have similar mean magnification factors. We can then measure \xp\ directly in the image plane without the need for the magnification correction. We discuss how the assumption can be violated, especially in the extreme case of \sous, in \S~\ref{ss:dis-Xf}.  We empirically measure the \xp\ factor in such individual galaxies, as well as find evidence for AGNs, to provide an important check on how the locally calibrated $X_{\sc HMXB,r}$ may be applied to galaxies with the most extreme SFRs observed exclusively at high-$z$ and to offer insights into high-energy processes in \sous.

\begin{table*}
\begin{center}
\caption{XMM-Newton Observation Log}
\label{t:obs}
\begin{tabular}{lccr}
\hline\hline 
Target & OBSID$^a$ & OBS Date Period & MOS1/MOS2/pn Exp$^b$ \\
Name & & (yyyy-mm-dd/yyyy-mm-dd) & (ks/ks/ks)\\
\hline 
\xsa & 0882720(301,601) & 2021-05-20/2021-07-03 &169.4/168.9/133.5\\
\xsb & 0882720(101,401) & 2021-12-26/2022-01-12 &134.5/134.6/95.21\\
\xsc & 0882720(201,501,701,901) &2021-10-31/2021-12-02 &212.1/212.1/172.2\\
\hline		
\end{tabular} 
\end{center}
{$^a$ Individual exposure numbers are enclosed in the parentheses.
$^b$ Background flare-cleaned exposures of the individual instruments.
}
\end{table*}

The organization of the rest of the paper is as follows. 
We describe our selection of targets (including their key multi-wavelength properties) in \S~\ref{s:targ}, the \xmm\ observations and data analysis in \S~\ref{s:data}, and present the results in \S~\ref{s:res}. In \S~\ref{s:dis} we compare the results with those obtained for local star-forming galaxies, including (U)LIRGs, and discuss the origin of the enhanced \xp\ and the implications for the detection of the AGNs observed in our targets. \S~\ref{s:sum} gives a summary of our results, conclusions, and near-future prospects. Additional multi-wavelength data reduction and modeling are detailed in Appendices \ref{a:sma},  \ref{a:PJ105322-IRAC}, \ref{a:PJ1053-lens-model} and \ref{a:lens-X-ray}.  All images presented in this paper have north up and east to the left. We assume the $\Lambda$CDM cosmology\footnote{\url{https://astro.ucla.edu/~wright/CosmoCalc.html}} 
with {H\textsubscript{o}} = 69.6 ${\rm~km~s^{-1}~Mpc^{-1}}$, {$\Omega$\textsubscript{M}} = 0.286 and {$\Omega$\textsubscript{$\Lambda$}} = 0.714.

\section{Target description}\label{s:targ}

\begin{figure}
\centerline{
\includegraphics[width=1.0\linewidth,angle=0]{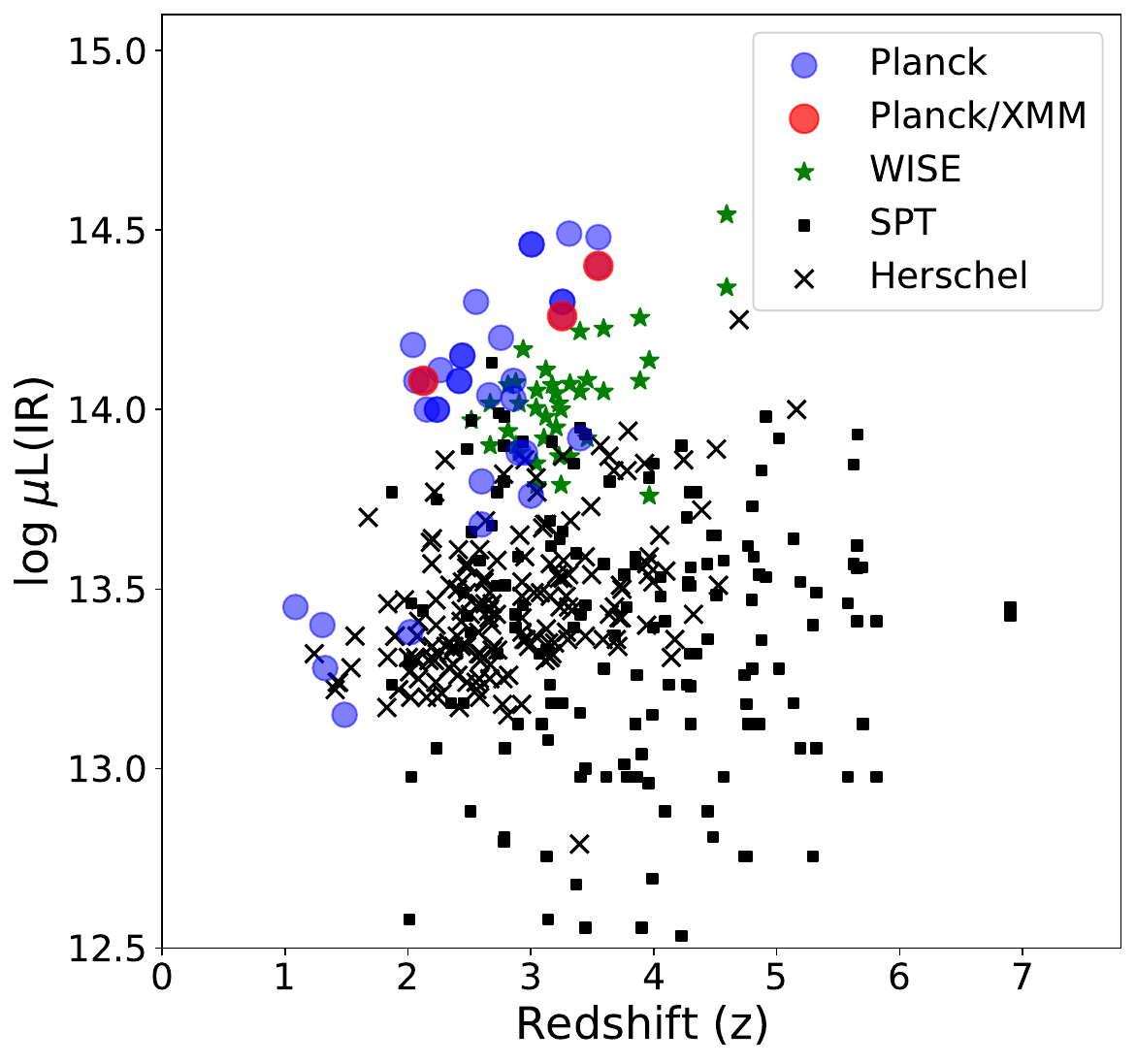}
}\caption{Apparent IR ($8-1000~\mu$m) luminosity (without magnification correction) vs. redshift plot of DSFGs, adapted from \citet[][see also \citet{Harrington2016}]{Berman2022}, but with the three \ins\ \sou\ targets highlighted here with filled red circles. This plot mainly shows the PASSAGES DSFGs \citep[blue dots][]{Harrington2016,Berman2022}, compared with other sources identified by  the WISE  \citep{Tsai2015}, South Pole Telescope (SPT) \citep{Vieira2013,Weiss2013,Reuter2020}, and Herschel \citep{Bussmann2013,Wardlow2013,Bakx2018,Bakx2020,Urquhart2022,Cox2023}. 
}
\label{f:f2}
\end{figure}
Our targets for the \ins\ observations (Table~\ref{t:targets}; Figs. \,\ref{f:f2}-\ref{f:f5}) are selected from the \planck\ All-Sky Survey to Analyze Gravitationally-lensed Extreme Starbursts (PASSAGES). This survey provides a sample of 30 gravitationally-lensed DSFGs representing the brightest IR galaxies observed in the Universe \citep[Fig.~\ref{f:f2}][]{Canameras2015,Harrington2016,Frye2019,Trombetti2021,Berman2022,Kamieneski2023}. A wealth of (sub)arcsecond multi-wavelength data have been taken for these DSFGs, using facilities including the Hubble Space Telescope (\hst), the Submillimeter Array (SMA), the Atacama Large Millimeter/submillimeter Array (ALMA), the Northern Extended Millimeter Array (NOEMA), 
and the Karl G. Jansky Very Large Array (VLA). These data have helped us to probe various stellar and interstellar components of the underlying DSFGs. 

Although our intended measurements as presented here are largely based on the observed (apparent) emission from the lensed DSFGs, the ultimate understanding of the underlying astrophysics will still be related to their intrinsic properties. Detailed lens modeling has been performed for many of the PASSAGES DSFGs \citep[e.g.][]{Frye2019, Frye2023a,Kamieneski2023}, showing that most of them are moderately to highly magnified by $\mu = 2 - 28$ \citep[see also][]{Bussmann2013,Geach2018,Canameras2018}. 
The lensing magnification depends on the location and hence the intensity distribution across a galaxy at a  wavelength. The quoted mean magnification factors for our targets in Table~\ref{t:targets} are best estimated with the ALMA CO(J=3-2) emission data for \xsb\ (comparable to that with the dust continuum emission and close to $\mu \approx 16$ with the \hst\ 1.6-$\mu$m data; \citet{Liu2023}), with the VLA 6 GHz continuum data for \xsa\  \citep{Kamieneski2023}, and with the SMA 850 \micron\
 data for \xsc\ (Appendix~\ref{a:PJ1053-lens-model}).
Our targets are mostly {\sl intrinsically luminous} [$L_{IR} \approx 1 \times10^{13} {\rm~L_{\odot}}$], hence \sous, rivaling the luminosities of even the brightest known unlensed objects at their respective redshifts. De-lensed images show that SF in the DSFGs tends to be widely distributed with the intrinsic sizes of far-infrared continuum regions ($R_e = 1.7 - 4.3$~kpc; median 3.0~kpc; \citet[][ and references therein]{Kamieneski2023}). In contrast, low-$z$ ULIRGs, which tend to form stars in compact nuclear regions, are often unresolved and have their core sizes $\lesssim 1.5$~kpc \citep[e.g.,][]{Diaz-Santos2010}. 

Our main selection criteria for the \xmm\ targets are as follows: 1) high apparent 1.1-mm flux $\gtrsim 60$~mJy and SED-inferred far-IR luminosity $\gtrsim 10^{14} L_\odot$ (e.g., Fig.~\ref{f:f2}) to maximize detection of the X-ray emission associated with extreme SF; 2) no evidence for AGN in the radio-infrared spectral energy distribution (SED) analysis \citep{Berman2022} to minimize its potential contamination, and 3) lensing mainly by a single galaxy or a small group of galaxies without any evidence for AGN (e.g., radio jets) to minimize X-ray contamination from them. Nevertheless, our selected three targets can still be complex systems with multiple galaxy components. We describe the main characteristics of the targets in Table~\ref{t:targets} and below:

\begin{figure*}
\centerline{
\includegraphics[width=1.0\linewidth,angle=0]{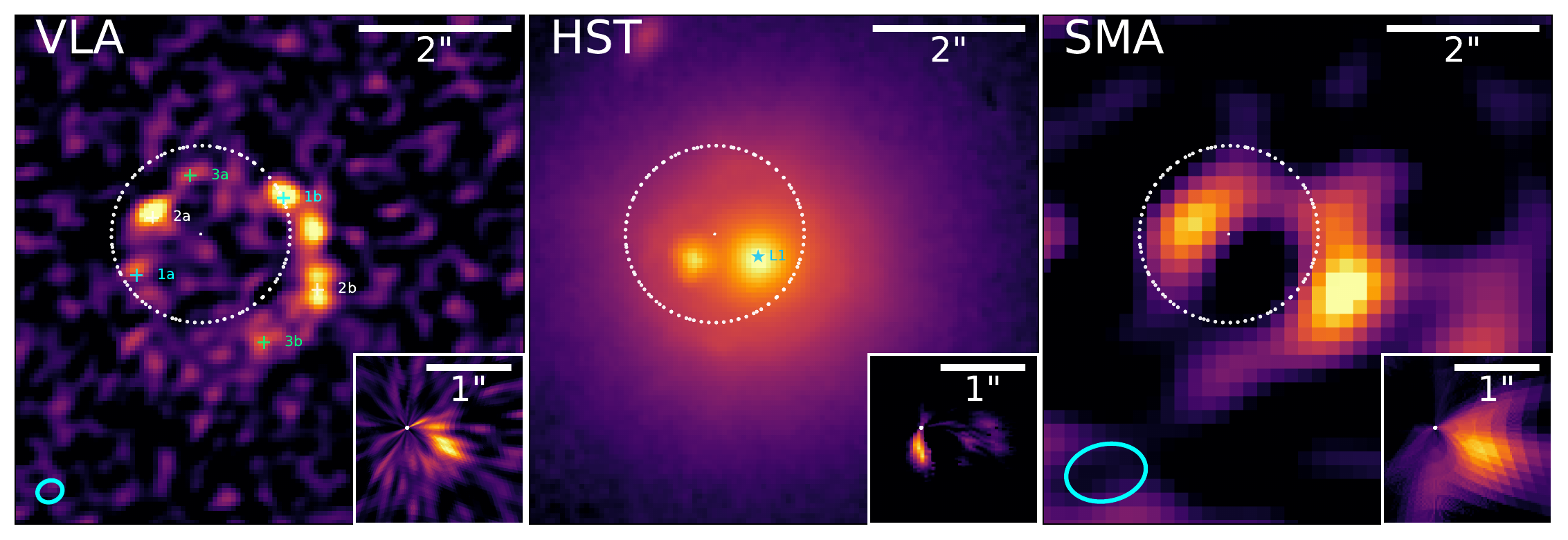}
}\caption{Multi-wavelength montage of the \xsa\ field, showing VLA 6 GHz continuum, \hst/WFC3 F160W, and SMA 1.2-mm continuum. Zoomed-in insets show the reconstructed source-plane structure. Light from the foreground lensing galaxy is subtracted before reconstructing the \hst\ image. White dotted curves show the model-derived caustic and critical curves from \citet{Kamieneski2023}. The synthesized beams for interferometric observations are shown in cyan in the lower left. Identified image systems used in the lens modeling are labeled on the image where they are most prominent.
}
\label{f:f3}
\end{figure*}
\begin{figure*}
\centerline{
\includegraphics[width=1.0\linewidth,angle=0]{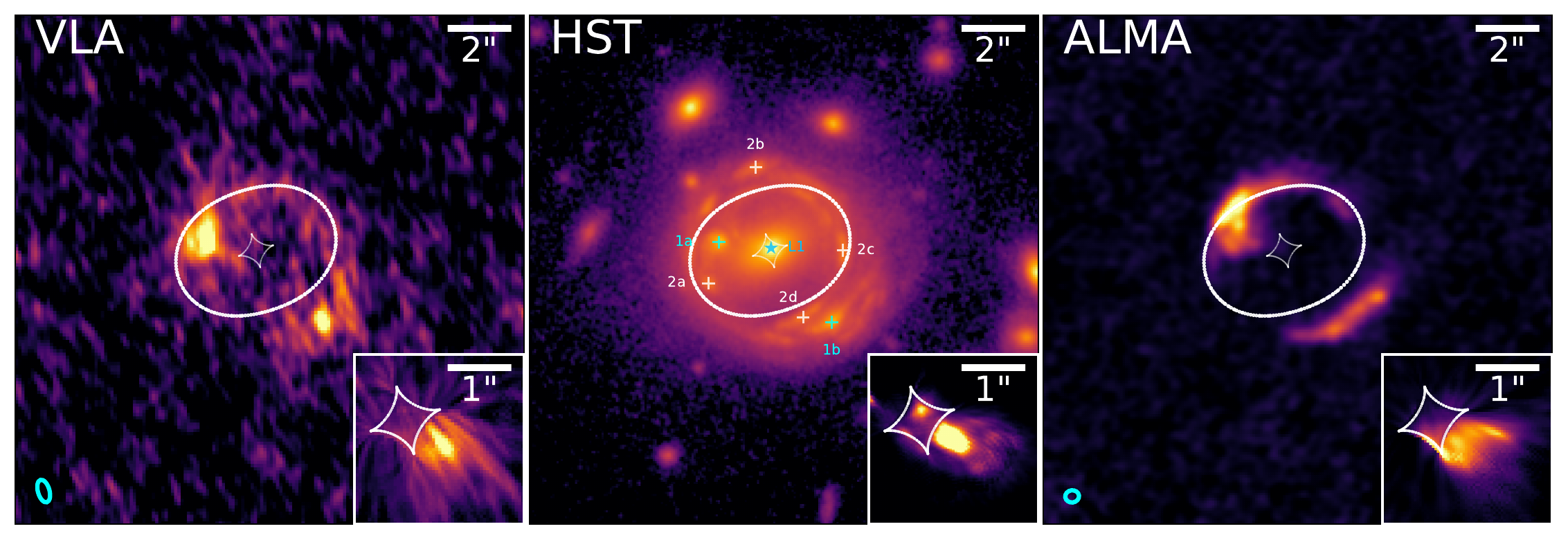}
}
\caption{
Multi-wavelength montage of the \xsb\ field, as with Figure \ref{f:f3}, but with the ALMA 1.1-mm continuum image. 
}
\label{f:f4}
\end{figure*}

\begin{figure*}
\centerline{
\includegraphics[width=1.0\linewidth,angle=0]{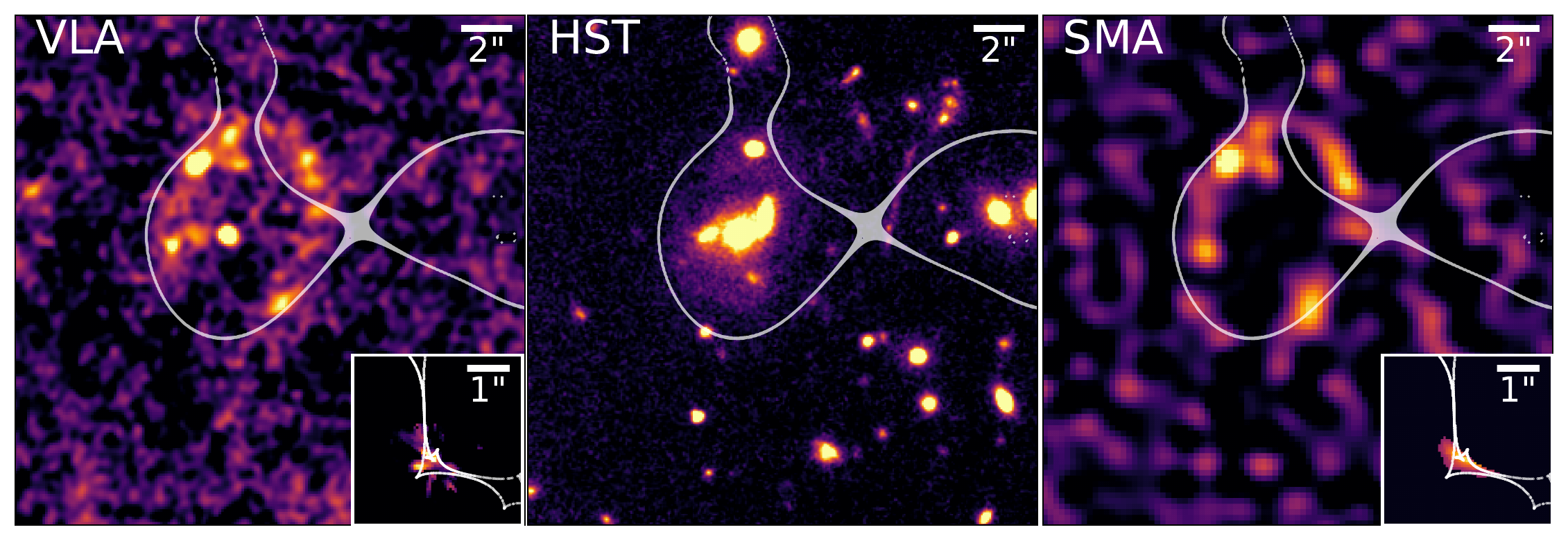}}
\caption{Multi-wavelength montage of the \xsc\ field, as with Fig.~\ref{f:f3}, together with the critical curve of the strong lensing (see Appendix~\ref{a:PJ1053-lens-model}). 
No clear counterpart of the DSFG is detected in the \hst/F160W image, but lensed features east of the central group seen in the image, together with the radio/sub-mm data, are used to constrain the lens model. Hence the source-plane reconstruction is only performed for radio/sub-mm data.
}
\label{f:f5}
\end{figure*}

\noindent {\bf \xsa\ (or G104.43+66.26)}: 
This \sou\ has the smallest Einstein radius (2$^{\prime\prime}$ diameter) and lensing amplification among our targets (Table~\ref{t:targets}). The lensed image of the galaxy is most vividly seen as a nearly complete ring in the VLA 6 GHz continuum and consists of several emission clumps larger than the synthesized beam (Fig.~\ref{f:f3}). Three doubly imaged families of these radio clumps are identified and labeled in the left panel of Fig.~\ref{f:f3}. These clumps are used to construct the lens model of the source \citep{Kamieneski2023}. The two relatively bright and unlabeled clumps between '1b' and '2b' are not used in the lens modeling; their counterparts on the east side of the ring are not clear and may be confused with the unresolved peaks '1a' and '2a'. 
The ring of the DSFG is also seen in the SMA 1.2-mm continuum image at $\sim$0.8$^{\prime\prime}$ resolution (Fig.~\ref{f:f3} right panel), further confirming the lensed star-forming regions (see Appendix~\ref{a:sma}). The lensing is dominated by a foreground compact galaxy labeled as 'L1' in the {\sl HST}/WFC3 1.6 \micron\ image (Fig.~\ref{f:f3} middle panel), which shows a faint counterpart of the lensed DSFG ring, more apparent after the foreground light is subtracted with GALFIT  \citep{Kamieneski2023,Lowenthal2023}. However, this counterpart is quantitatively very uncertain. As a result, the re-constructed source-plane image in the 1.6 \micron\ band may be problematic and is presented here just for completeness. 

\noindent {\bf \xsb\ (or G190.40-83.77)}: This \sou\ has the largest apparent SFR among our targets \citep[Table~\ref{t:targets};][]{Berman2022,Liu2023}. Its {\sl HST}/WFC3 1.6 \micron\ image shows a nearly complete double Einstein ring with an average diameter of 6$^{\prime\prime}$, while the VLA 6 GHz and the ALMA 1.1-mm continuum and CO images show two major resolved arcs, which lie mostly in the middle of the ring  \citep[Fig.~\ref{f:f4};][]{Kamieneski2023}. The results from detailed modeling of the VLT/ERIS near-IR IFU data, as well as the ALMA CO data, have been reported by \citet{Liu2023}. The detection of the H$\alpha$, H$\beta$, [NII], and [SII] lines shows a slightly supersolar gas-phase metallicity and a high mean Balmer decrement (observed H$\alpha$/H$\beta$ flux ratio of $9.2 \pm1.2$, corresponding to $A_V \approx 3$). However, the stellar continuum spectrum suggests a lower value ($A_V \approx 1.5$), indicating that the attenuation is highly structured.  The galaxy is well resolved in high resolution ($\sim$ 0\farcs2) ALMA continuum and CO observations, after correction for lensing, revealing a secularly evolving, massive galaxy ecosystem with, unlike most known \sous, no major merger signatures. The systematic rotation in both cold and ionized gas tracers is consistent with a massive baryonic disk ($M_{\rm baryon}\gtrsim 10^{11.1} {\rm~M_{\odot}}$) living in a massive dark matter halo ($M_{\rm DM}\sim 10^{13.2} {\rm~M_{\odot}}$). The massive disk of \sou\ forms stars at an intrinsic rate of $\sim 1.5 \times 10^3 {\rm~M_{\odot}~yr^{-1}}$ and has a S\'{e}rsic profile half-light radius of $\sim 2.9$~kpc. The disk also shows indications of two grand spiral arms. The compact bluer peak seen within the critical curve in the inset forms the four images labeled 2a, 2b, 2c, and 2d in the middle panel of Fig.~\ref{f:f4}. By measuring the half-light radius of these images as $\sim 4^{\prime\prime}$ in the image plane and dividing by the linear magnification factor of $\mu^{0.5}$, we estimate the effective radius of the peak as $\sim 0.9 - 1.3$~kpc in the source plane. Accounting for about 1/7 of the rest-frame g-band flux of the \sou, the peak likely represents a giant star-forming region -- a typical signature expected for high Toomre instability in massive high-$z$ disks. In contrast, the stellar bulge of the galaxy is largely devoid of cold gas, suggesting inside-out quenching.

\noindent {\bf \xsc\ (or G145.25+50.84)}: This \sou\ (Fig.~\ref{f:f5}) has the highest redshift and has the largest gravitational amplification among our \xmm\ targets (Table~\ref{t:targets}).  
It is one of the PASSAGES sources also studied with  {\sl Herschel}  \citep{Harrington2016,Berman2022}, independently identified by \citet{Canameras2015}, and followed-up with \hst\ observations \citep{Frye2019}. Though not particularly bright, intrinsically,  the source shows highly excited CO emission and very broad line profiles \citep[e.g.,][]{Harrington2021}. The SMA $\sim$200-230 GHz continuum image of the source (see Appendix~\ref{a:sma}) shows an almost complete elongated Einstein ring with an average diameter of 11$^{\prime\prime}$ \citep[see also ][]{Canameras2015}. 
The ring is also well detected in the {\it Spitzer}/IRAC 3.6 and 4.5 \micron\ imaging data (Appendix~\ref{a:PJ105322-IRAC}).
In contrast, the \hst\ 1.6-\micron\ image shows little evidence for the ring, indicating strong dust attenuation. 
The \hst\ 1.6-\micron\ image, however, reveals additional lensed features from galaxies in the background other than the \sou. These features are included in the lens modeling detailed in Appendix~\ref{a:PJ1053-lens-model}. The result is illustrated in Fig.~\ref{f:f5}. 

\section{\xmm\ Observations and Data Analysis}\label{s:data}

\subsection{\xmm\ observations}\label{ss:d-obs}
The relevant data sets of the \xmm\ observations are obtained from a set of three X-ray CCD cameras comprising the European Photon Imaging Camera (EPIC) \footnote{\url{https://www.cosmos.esa.int/web/xmm-newton/technical-details-epic}}. Two of the cameras are MOS (Metal Oxide Semiconductor) CCD arrays, while the other uses pn CCDs and is referred to as the pn camera. The angular resolution of the EPIC pn is about 6.6$^{\prime\prime}$ (FWHM) or 15$^{\prime\prime}$ (80\% energy-encircled radius), while the MOS detectors have slightly better resolutions (e.g. 6$^{\prime\prime}$ FWHM).
 Table~\ref{t:obs} summarizes the key parameters of the \ins\ observations. Limited by the orbital period, each observation of a target was split into multiple exposures. For this study, we use data from the European Photon Imaging Camera (EPIC) instruments, MOS1/2 plus pn cameras.

 \subsection{Imaging Data}\label{ss:d-imaging}
 
\begin{figure}
\centerline{
\includegraphics[width=1.0\linewidth,angle=0]{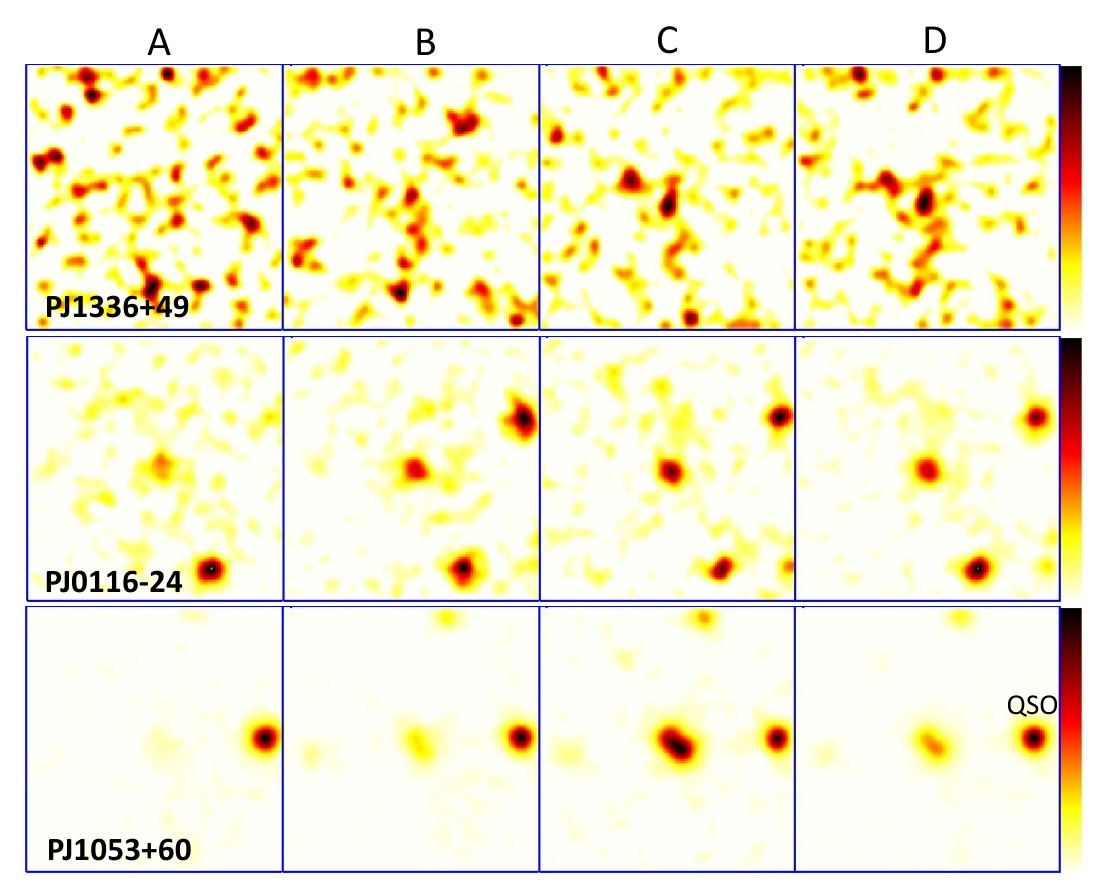}
}
\caption{
Combined \ins\ (MOS1/2+pn) count intensity images of our target \sous\ ($1.5' \times 1.5'$ field) in the 0.3-0.7~keV (Col. A), 0.7-1.2~keV (B), 1.2-7.0~keV (C), and 0.3-7~keV (D). The images are all centered on the respective targets and are smoothed with a Gaussian with FWHM=6''. The white-black color range is scaled from the median to the maximum intensities of each image. As a point-like source reference, a known QSO (SDSS J105315.15+605145.7) at z=1.1908 is marked in the field of \xsc.
}
\label{f:f6}
\end{figure}
\begin{figure}
\centerline{
\includegraphics[width=1\linewidth,angle=0]{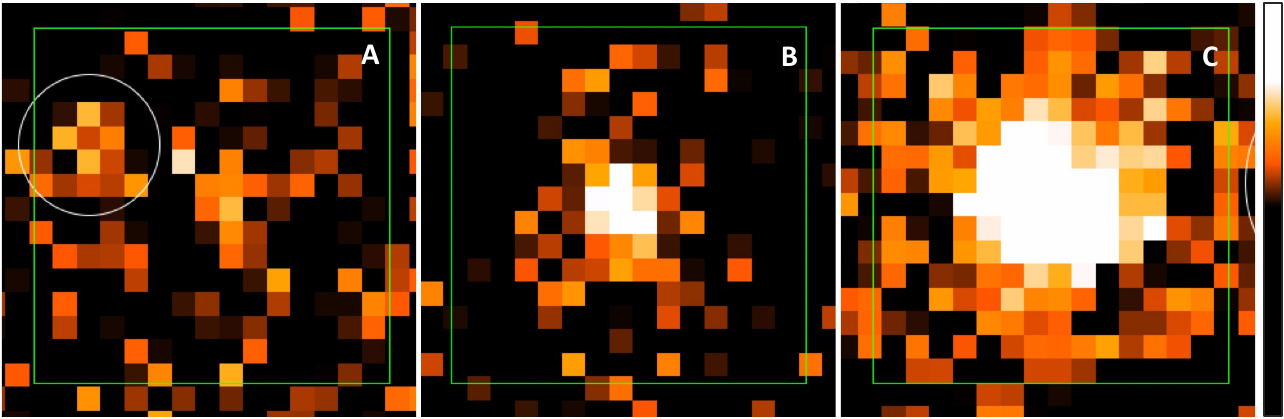}
}
\caption{
Unsmoothed combined pn 0.5-4.5~keV count images for the three targets: 
{\bf (A)} \xsa,  {\bf (B)} \xsb, and  {\bf (C)} \xsc. The white circle in (A) marks the region that is excluded for the construction of the radial intensity profile shown in Fig.~\ref{f:f8}A. The green box in each panel shows the $1' \times 1'$ field of view. 
}
\label{f:f7}
\end{figure}

\begin{figure*}
\centerline{
\includegraphics[width=1.0\linewidth,angle=0]{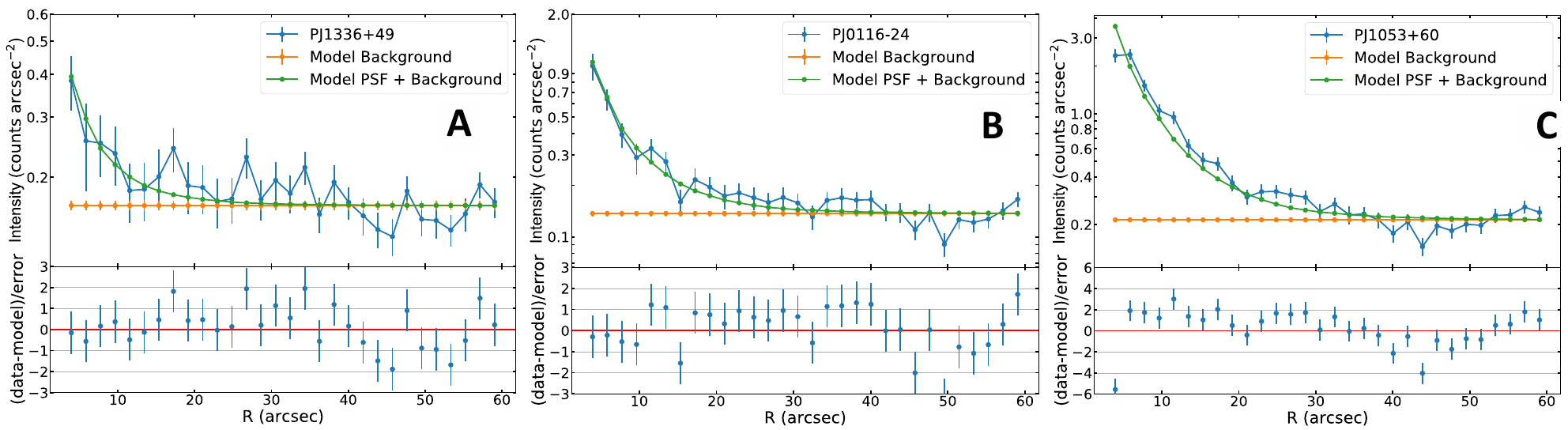}
}
\caption{
Same as Fig.~\ref{f:f9}, but for
 our targets: {\bf (A)} \xsa,  {\bf (B)} \xsb, and  {\bf (C)} \xsc\ (see also Table~\ref{t:sbp}).
}
\label{f:f8}
\end{figure*}

\begin{figure}
\centerline{
\includegraphics[width=1.0\linewidth,angle=0]{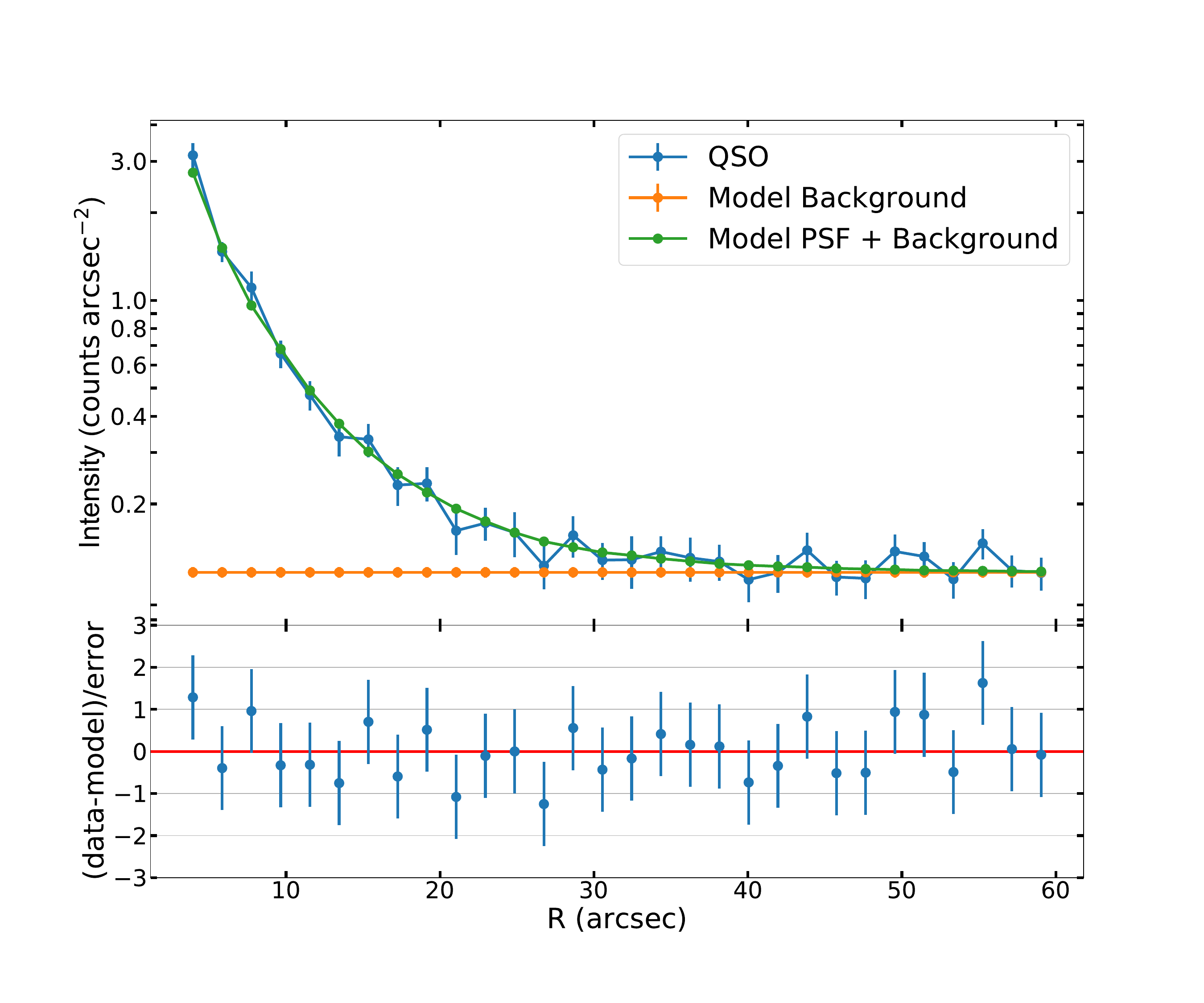}
}
\caption{Comparison of the PSF with the radial intensity profiles of the pn 0.5-4.5~keV emission for QSO SDSS J105315.15+605145.7 (see also Table~\ref{t:sbp}).
}
\label{f:f9}
\end{figure}

We use the pipeline product (PPS) directly to visualize the data.
To obtain maximum counting statistics of the imaging data, we combine the PPS MOS1/2+pn count intensity images produced for each exposure in the 0.3-0.7~keV, 0.7-1.2~keV, and 1.2-7.0~keV bands. According to the XMM-Newton specifications (XMM-SOC-GEN-ICDD-0024, Issue 4.6), these PPS images have been corrected for vignetting and exposure, while events from MOS1/2 cameras are given weights 5.41, 4.64, and 2.96 in the three bands for the pn ones, from which minimum instrument backgrounds and out-of-time events have also been removed. We further merge these combined images of individual exposures, excluding pixels labeled NaN, accounting for the slightly different pointing directions, and using the PPS (band 8000) exposure images as weights. The merged images in the three bands are finally added together to form a broadband (0.3-7~keV) image of the target (Fig.~\ref{f:f6}).

We also perform a preliminary spatial analysis of the \xmm\ data mainly to check how point-like the X-ray emission is for each target (Table~\ref{t:sbp}). Fig.~\ref{f:f7} shows the close-up pn count images of the three targets in the 0.5-4.5~keV band. The signal to noise ratios ($S/N$) in these images tend to be optimal. The imaging data of \xsa\ look quite noisy, although there is obviously a net excess of X-ray emission. This is more apparent in the 1-D radial intensity profile of the emission shown in Fig.~\ref{f:f8}A. The other two targets are much brighter. We fit the profile of each target to obtain the normalization of the point spread function (PSF) and the local constant background. The PSF is approximately constructed at an effective photon energy of 3 keV and is insensitive to a small change in this energy. We check this procedure with SDSS J105315.15+605145.7 - a QSO at $z =1.1908$, 58$^{\prime\prime}$ west of \xsc\ (Fig.~\ref{f:f6}). This source is quite bright and is expected to be point-like, as confirmed by the comparison shown in Fig.~\ref{f:f9} and Table~\ref{t:sbp}, suggesting that our adopted PSF is good. Fig.~\ref{f:f8} presents the comparison of the profiles with the best-fit normalizations and local backgrounds of the targets. The profile of \xsc\ is clearly inconsistent with being point-like at a high confidence level (Table~\ref{t:sbp}). Based on the present analysis of the X-ray data, we cannot reject the point-like nature of the emission for both \xsb\ and \xsa\ with $\gtrsim 80\%$ confidence. However, a close examination of the image of \xsb, as well as a preliminary 2-D fit, suggests that this target may not be point-like in the 2-D emission distribution (e.g., Fig.~\ref{f:f10}A). While more careful spatial analysis is required to make a firm conclusion, we focus here on the spectral analysis and results.

\begin{table}
\begin{center}
\caption{PSF comparison results}
\label{t:sbp}
\begin{tabular}{lcr}
\hline\hline 
Target & $\chi^2$/d.o.f & P-value$^a$\\
\hline 
\xsa & 29.7/29 & 0.43\\
\xsb & 36.6/29 & 0.15\\
\xsc & 98.2/29 & $1.9 \times 10^{-9}$\\
QSO SDSS J105315.15+605145.7 & 14.5/29 & 0.99\\
\hline		
\end{tabular} 
\end{center}
{$^a$ The probability for a source to be point-like.
}
\end{table}

\subsection{Spectral Data}\label{ss:d-spec}
For spectral analysis, we independently process the \ins\ data with the Science Analysis System (SAS), primarily for our X-ray spectral analysis of each target. This process includes creating event files separately for the MOS and pn data using the \texttt{emproc} and \texttt{epproc} routines; filtering out time intervals with high background activity\footnote{\url{https://www.cosmos.esa.int/web/xmm-newton/sas-thread-epic-filterbackground}}; and applying the \texttt{edetect chain} to detect discrete sources. We also use \texttt{evselect} to extract the MOS1/2 and pn spectra and their corresponding local background spectra, and \texttt{rmfgen} and \texttt{arfgen} to obtain the corresponding instrument response and effective area files. Spectra from the same instrument but different exposures of the same target are combined with \texttt{epicspeccombine}. The two MOS spectra are then combined due to their similar properties. 

The centers of the source extraction regions are determined from the \texttt{edetect chain} source detection results. The X-ray centroid coordinates (R.A./Dec. J2000) of \xsb, \xsa, and  \xsc\ are 1:16:46.63/-24:37:03.0, 13:36:35.13/49:13:12.9, and 10:53:22.08/60:51:43.2, respectively. The on-source spectral extraction aperture of each source is 20$^{\prime\prime}$ radius with a corresponding annulus background extraction region with an inner radius of 35$^{\prime\prime}$ and an outer radius of 175$^{\prime\prime}$. The aperture size is large enough to obtain the majority of the spectral data, while regions are excluded for other detected X-ray sources in each background annulus.

\subsection{Spectral Modeling}\label{ss:d-spec-modeling}

The modeling of the background subtracted spectral data must account for potential X-ray contributions of several types: (discrete and diffuse) X-ray sources in the foreground lenses, and the AGN/non-AGN components of the lensed \sous. In Appendix~\ref{a:lens-X-ray} we estimate the foreground lens contributions and find them to be negligible (count rates $\lesssim 1\%$ in the 0.5-2~keV band and less in broader bands). Therefore, we focus here on modeling contributions from the \sous. 
 
Our spectral modeling uses a combination of the X-ray spectral fitting package \texttt{Xspec} \citep{Arnaud1996} and the Bayesian X-ray Analysis (\texttt{BXA}) software \citep{Buchner2014}\footnote{https://johannesbuchner.github.io/BXA/index.html}. We further use the python package,
\texttt{chainconsumer} \citep{Hinton2016} to improve the presentation of the corner plots. The \texttt{BXA} takes the initial best-fit parameters from \texttt{Xspec} to perform the MCMC analysis. We use the High Energy Astrophysics Science Archive Research Center (HEASARC) routine \texttt{ftgrouppha} to group spectral channels of the resulting combined MOS and pn spectra to obtain certain counting statistics. For the use of the C-statistic (called cstat in \texttt{Xspec}) or more precisely the W-statistic (wstat), appropriate for dealing with the Poisson data (\url{https://heasarc.gsfc.nasa.gov/xanadu/Xspec/manual/node319.html}), we group the spectral data to obtain a minimum of seven counts per bin, as recommended in \texttt{Xspec} (\url{https://asd.gsfc.nasa.gov/Xspecwiki/low_count_spectra}).
For ease of the visualization of the spectra, we group them to obtain background-subtracted $S/N > 1$ per bin in \texttt{Xspec}. 

The sophistication of our spectral modeling is limited by the counting statistics of the data. We start with simple models with the minimum number of fitting parameters and add complications only when justified, either statistically or physically. All spectral components are subject to the Galactic foreground absorption, modeled with \texttt{tbabs} with column density fixed at $N_{H,G}$ (Table~\ref{t:targets}). We fit an effective intrinsic (rest-frame) X-ray absorbing gas column density ($N_{H,HMXB}$) with the \texttt{Xspec} model \texttt{ztbabs}. The \texttt{Xspec} default solar metal abundances are assumed for all absorbing gases. 
We model the intrinsic HMXB contribution with a redshifted power-law \texttt{zpowerlw}; the corresponding \texttt{Xspec} model syntax is \texttt{const*clumin(zpowerlw)}, where the convolution model \texttt{clumin}, allows us to conveniently pre-set the intrinsic {\sl rest-frame} luminosity of the contribution to a reference luminosity (which we set to be $X_{\sc HMXB,r}   \times$ 
 the SFR of a target), while the model \texttt{const} or \xp\ serves as the normalization of the component to be fitted. Taken together, we have a basic set of spectral models:
\noindent {\texttt{tbabs * (ztbabs * const*clumin(zpowerlw))}}.
We find that this model is sufficient for the characterization of \xsa, where the three free parameters to be constrained by the spectral fit are: \xp, the photon index of the power law ($\Gamma_{HMXB}$ or $\Gamma_{nonAGN}$), and the intrinsic X-ray absorbing column density ($N_{H,HMXB}$ or $N_{H,PJ1336}$ used later in the two-source joint fit). 

However, \xsb\ and \xsc\ show X-ray evidence for the presence of AGN (see \S~\ref{s:res}). Therefore, we have included another redshifted power-law component, \texttt{ztbabs * clumin(zpowerlw)}, with the corresponding X-ray absorbing column density ($N_{H,AGN}$), AGN power-law photon index ($\Gamma_{AGN}$), and luminosity ($L_{X,AGN}$) to be fitted. 
We adopt a prior for the AGN photon index as a normal distribution with a mean of 1.95 and a standard deviation of 0.15
\citep[][see also \citet{Migliori2023,Zappacosta2023}]{Buchner2014}. 
From the fitted spectral models we derive the observed absorbed 0.5-8 keV flux for each spectral component ($f_{X,HMXB}$ or $f_{X,AGN}$). 

\begin{figure}
\centerline{
\includegraphics[width=1.1\linewidth,angle=0]{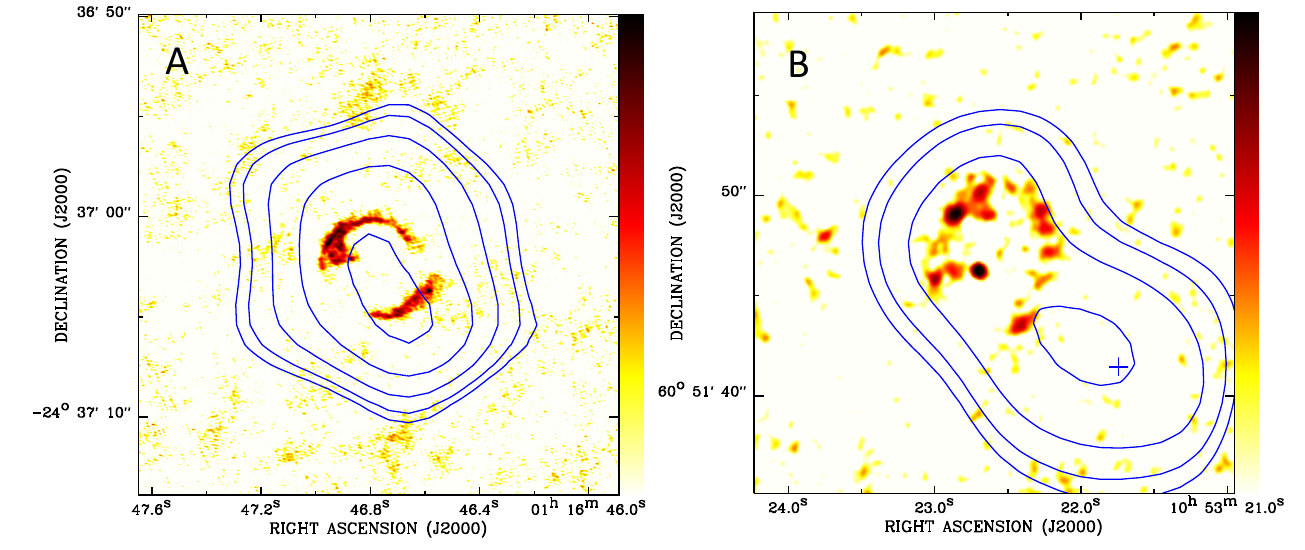}}
\caption{
\xmm\ 0.3-7~keV intensity contours overlaid on the (A) ALMA 1.1-mm image of \xsb\ and (B) VLA 6-GHz image of \xsc. These X-ray intensity contours (all in units of $10^{-2} {\rm~counts~s^{-1}~arcmin^{-2}}$) 
are at 0.4, 0. 6, 1, 1.6, and 2.4 above a local background of $\sim 1.4$ for \xsb\ and at 1, 2, 4, and 8 above a local background of $\sim 4.5$ for \xsc, while the X-ray images are constructed in the same way as for Fig~\ref{f:f6}, but smoothed with a Gaussian with FWHM$=3^{\prime\prime}$. The position of the AGN, identified in the \xsc\ field, is marked with a plus in panel B.
}
\label{f:f10}
\end{figure}

\section{Results}\label{s:res}

Fig.~\ref{f:f6} presents the merged MOS+pn images of our targets in the stacked broadband, as well as in individual energy bands. These images give an overview of the \xmm\ data quality of individual targets, with respect to their local environment. They are all clearly detected in the broadband images. Fig.~\ref{f:f7} shows the unsmoothed XMM pn data close-ups of the targets in the 0.5-4.5~keV band. It is clear that the \xmm\ counting statistics of \xsa\ is too limited to give useful 2-D spatial information. Fig.~\ref{f:f8} compares the radial intensity profiles of the targets with that of the PSF (Table~\ref{t:sbp}). We further present the close-ups of the smoothed broadband images of \xsb\ and \xsc\  in Fig.~\ref{f:f10}. The X-ray emission of \xsc\ is clearly not point-like, even in the radial profile analysis (Table~\ref{t:sbp}). In the smooth 2-D image, \xsb\ seems to be elongated, as indicated by the higher intensity contours, although this is yet to be quantified via detailed spatial analysis. The results from our spectral analysis are summarized in Table~\ref{t:spec}. In particular, our measured \xp\ values are compared with the reference calibration data in Fig.~\ref{f:f1}. In the following, we focus on describing the X-ray spatial and spectral properties of the targets.

\begin{table*}
\caption{Results of Spectral analyses}\label{t:spec}
\vspace{-0.2cm}
{
\begin{tabular}{lcccccccccr}
\hline\hline
Target & \xsa & \xsb & \xsc &\xsa/\xsb\\
\hline
Total \# of pn/2MOS counts & 464/399 & 536/601 & 4149/1702 \\
Net \# of pn/2MOS source counts & 50.2/50.8 & 219.3/135.3 & 1234.3/1003.4\\
\texttt{wstat}/d.o.f & 110.91/116 & 185.06/149 & 671.76/696 & 298.79/267 \\
Goodness (\%) & 21.2 & 42.4 & 90.6 & 48.0 \\
$\Gamma_{HMXB}$ & $2.79 (1.78,3.64)$ & $2.81 (2.05,4.75)$ & 2.29 (1.38,3.32) & $2.46 (2.09,3.32)$ \\
$X_{\sc HMXB} (3.9 \times 10^{39} {\rm~erg~s^{-1}}/{\rm M_\odot~yr^{-1}})$ & $4.2 (1.4,9.1)$& $3.7 (0.9,8.7)$ & $14.6 (8.7,23.8)$ & $3.4 (2.2,7.1)$\\
log[$N_{H,HMXB} (10^{22} {\rm~cm^{-2}}$)] & $-0.5 (-1.8,0.8)$ & $-0.3 (-1.8,3.4)$ & $-0.6 (-1.8,0.7)$ & $-0.5 (-1.8,0.7)$/$-0.8 (-1.9,2.1)$\\
-$f_{X,HMXB} (10^{-15} {\rm~erg~s^{-1}~cm^{-2}})$ & $0.88 (0.42,1.67)$ & $3.5 (0.1,6.6)$ & $4.7 (1.3,15.6)$ & $1.17 (0.59,1.68)$/$4.9 (1.6,6.4)$ \\
$\Gamma_{AGN}$ & .. & $1.99 (1.74,2.22)$ & $1.98 (1.79,2.19)$ & ../$1.96 (1.71,2.21)$\\
log[$N_{H,AGN} (10^{22} {\rm~cm^{-2}}$)] & .. & $-0.3 (-1.8,3.5)$ & $1.67 (1.52,1.81)$ & ../$0.6 (-1.8,3.6)$\\
log[$L_{X,AGN} ({\rm~erg~s^{-1}})$] & .. & $44.23 (42.33,44.56)$ & $45.85 (45.69,46.03)$ & ../$43.75 (42.23,44.54)$ \\
-$f_{X,AGN} (10^{-15} {\rm~erg~s^{-1}~cm^{-2}})$ & .. & $3.9 (0.1,7.1)$ & $35.9 (26.6,40.9)$ & ../$1.00 (0.05,6.06)$ \\
\hline		
\end{tabular} 
}
\vskip -0.05cm

\noindent{The goodness value of a fit represents the percentage of 1000 data simulations that have a smaller wstat/d.o.f., while the 90\% confidence interval of each parameter is included in the parenthesis next to the best-fit value. The unit of \xp\ is $X_{\sc HMXB,r}$ in Eq.~\ref{e:xsfr-r}. The AGN luminosity is calculated in the rest frame 0.5-8.0~keV band of each source, while the absorbed fluxes of the HMXB and AGN components ($f_{X,HMXB}$ and $f_{X,AGN}$) are inferred (hence the marked '-' ) from the best-fit model in the observed 0.5-8.0~keV band. 
}
\end{table*}

\begin{figure}
\centerline{
\includegraphics[width=0.95\linewidth,angle=0]{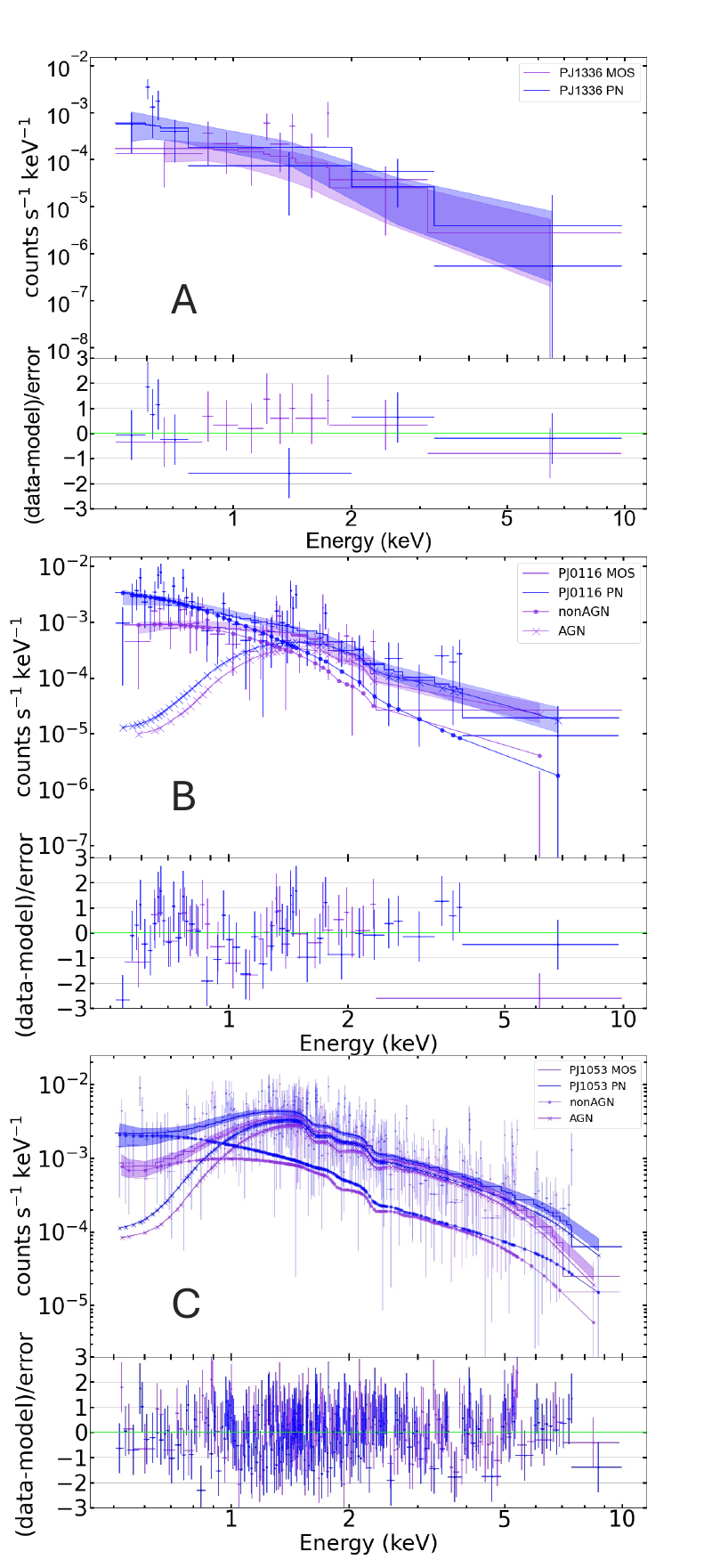}
}
\caption{
Spectral data and fitting models of individual targets: simple power law fit for \xsa\ (A); double power law fit for \xsb\ (B) and 
\xsc\ (C). The two components are also separately plotted in (B) and (C). 
}
\label{f:f11}
\end{figure}

\begin{figure}
\centerline{
\includegraphics[width=1.0\linewidth,angle=0]{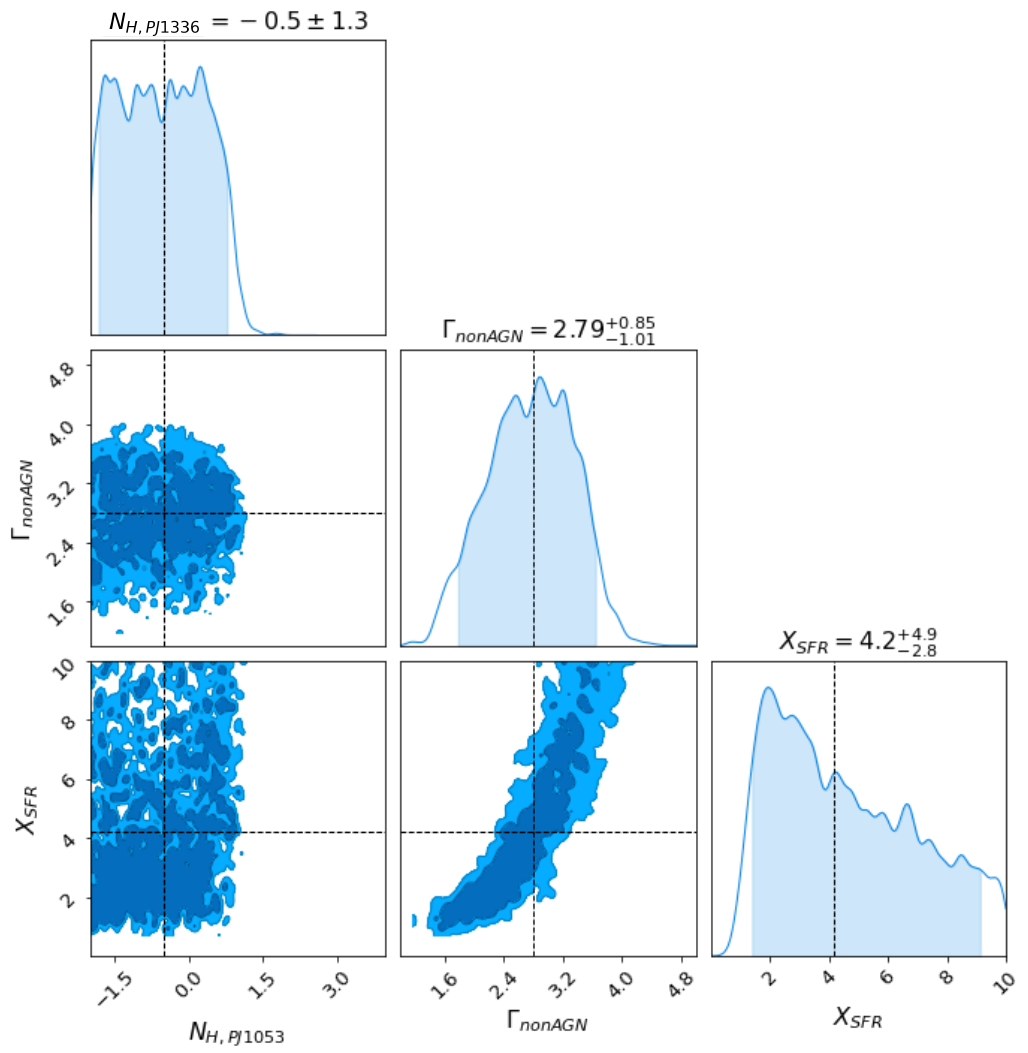}
}
\caption{
Corner plot of the model parameters from the fit to the spectrum of \xsa\ (Table~\ref{t:spec}). 
}
\label{f:f12}
\end{figure}

We start with \xsa\ -- probably the simplest case here. 
The detected net number of \xmm\ counts of the target is very limited (Fig.~\ref{f:f11}A; Table~\ref{t:spec}), which is expected if the X-ray emission from the \sou\ is dominated by HMXBs. Their spectrum can be well characterized by a power law with an intrinsic absorption column density of $\sim 3 \times 10^{21}$ cm$^{-2}$ (Figs.~\ref{f:f11}A and \ref{f:f12}; Table~\ref{t:spec}). The best-fit model gives \xp\ $\sim4$ or an apparent X-ray luminosity $L_X \approx 3 \times 10^{44} {\rm~erg~s^{-1}}$. The relatively steep spectrum (e.g., power law index $\Gamma_{HMXB} \sim 2.8$) is consistent with the hypothesis that the X-ray emission represents the collective contribution from the HMXBs of the galaxy. Although both the power law index and the absorption are not tightly constrained, we see no evidence for a significant AGN component, which should typically have a harder spectrum, consistent with the lack of a point-like source component in radio (Fig.~\ref{f:f3}A).

\begin{figure}
\centerline{
\includegraphics[width=1.0\linewidth,angle=0]{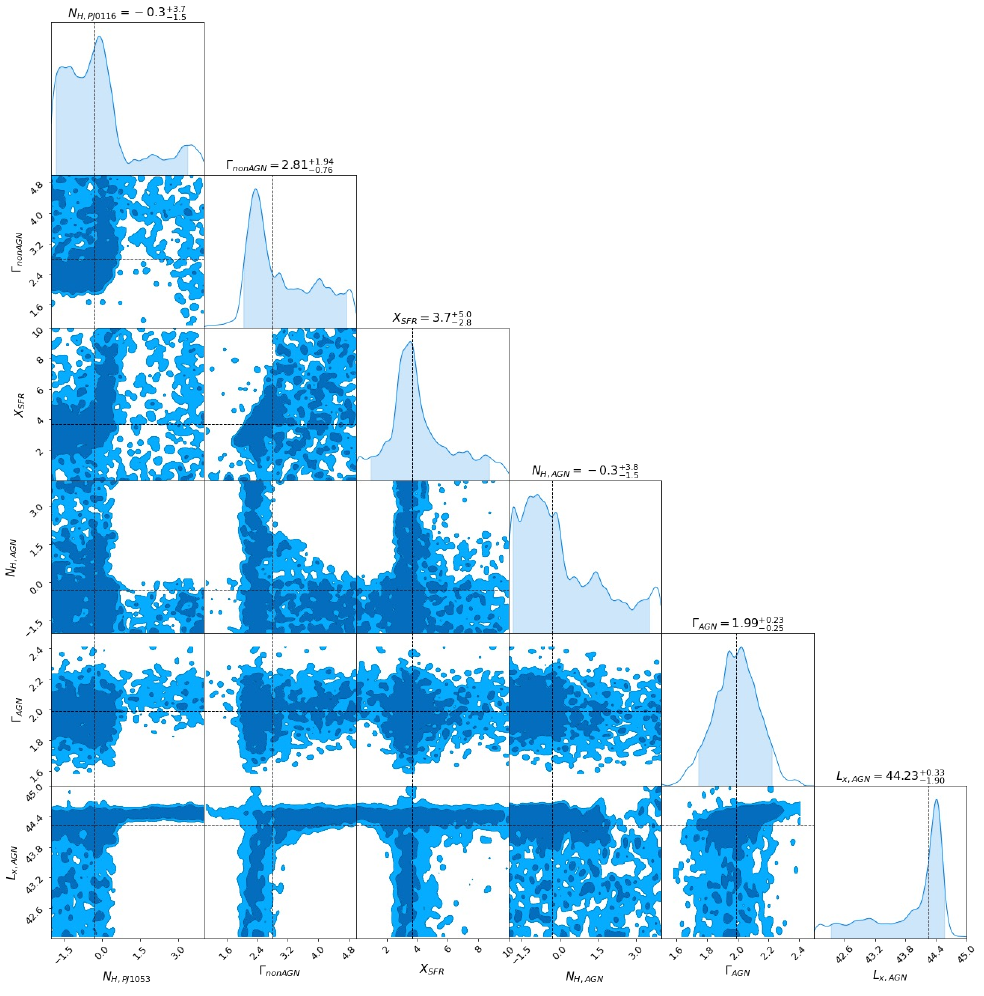}
}
\caption{
Corner plot of the model parameters from the fit to the spectrum of \xsb\ (Table~\ref{t:spec}). 
}
\label{f:f13}
\end{figure}

\begin{figure}
\centerline{
\includegraphics[width=1.0\linewidth,angle=0]{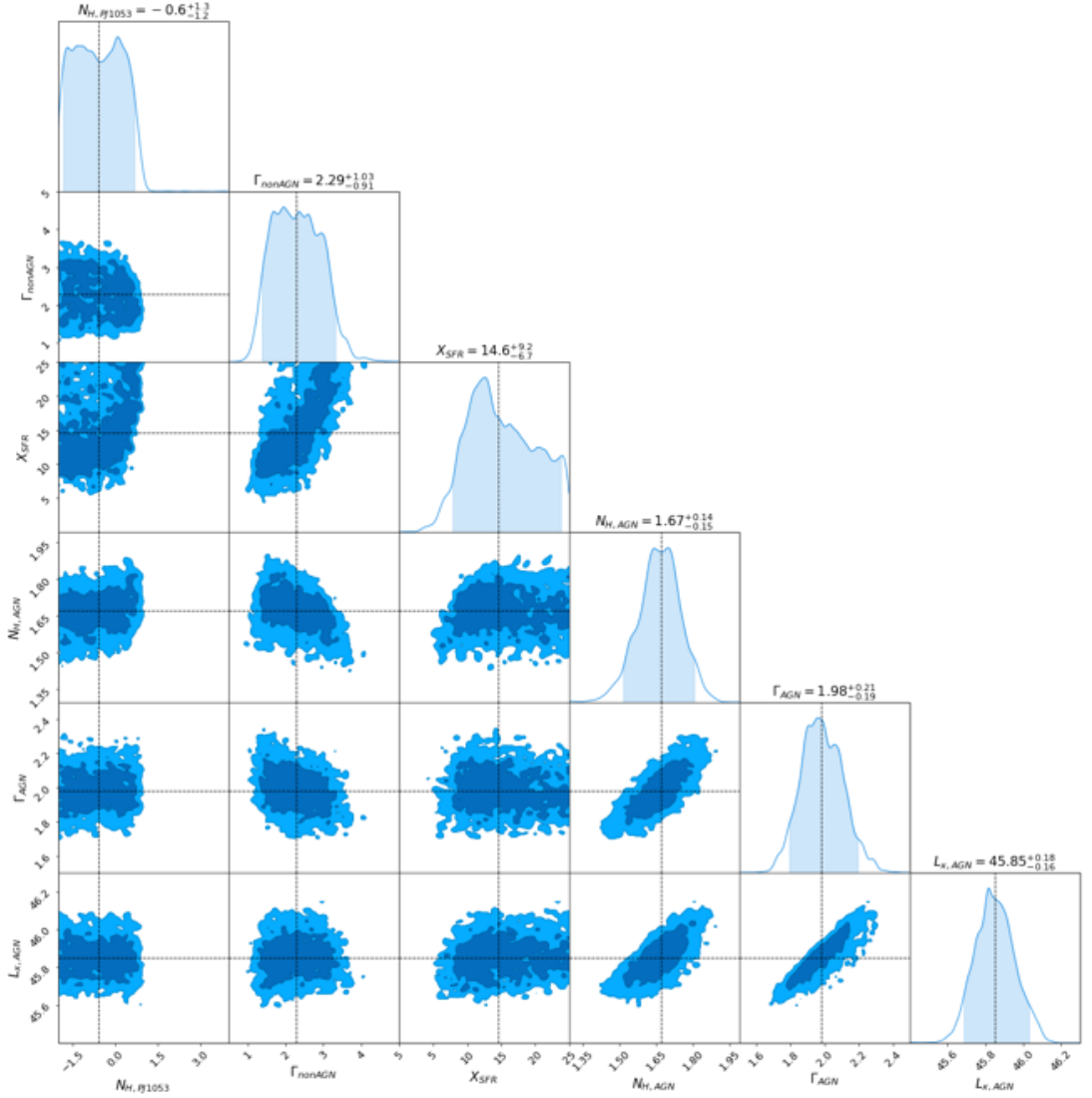}
}
\caption{
Corner plot of the model parameters from the fit to the spectrum of \xsc\ (Table~\ref{t:spec}). 
}
\label{f:f14}
\end{figure}

The spectra of \xsb\ and \xsc\ (Fig.~\ref{f:f11}B and C) do not fit as well with the simple power-law model with wstat/ d.o.f=189.65/152 (where d.o.f is the degree of freedom) and 817.94/699, respectively. We find that the double power-law model gives better fits to the spectra of these two sources (Fig.~\ref{f:f11}B-C, ~\ref{f:f13}, and ~\ref{f:f14}; Table~\ref{t:spec}). Their estimated luminosities are far too large to be consistent with the values expected from the \xp\ estimated from the above fit to \xsa\ or from the reference calibration $X_{\sc HMXB,r}  $ (Eq.~\ref{e:xsfr-r}), indicating a substantial AGN contribution. Indeed, a follow-up analysis of the data available in other wavelength bands supports this hypothesis (see \S~\ref{ss:dis-AGN}). Here, the fits with the double power-law model allow us to decompose these two contributions in \xsb\ and \xsc. Studies have shown that ULXs on average tend to have spectra that steepen substantially at rest frame energies of a few keV \citep[e.g.,][]{Bachetti2013,Brightman2018,Walton2020}, whereas an AGN spectrum intrinsically has a flatter power-law-like spectrum and tends to be flattened due to the Compton hump at rest frame energies $\gtrsim 10$~keV \citep[e.g.,][]{Reynolds1995,Brightman2018}. This makes a crude spectral decomposition of the two components possible.

\begin{figure}
\centerline{
\includegraphics[width=1.0\linewidth,angle=0]{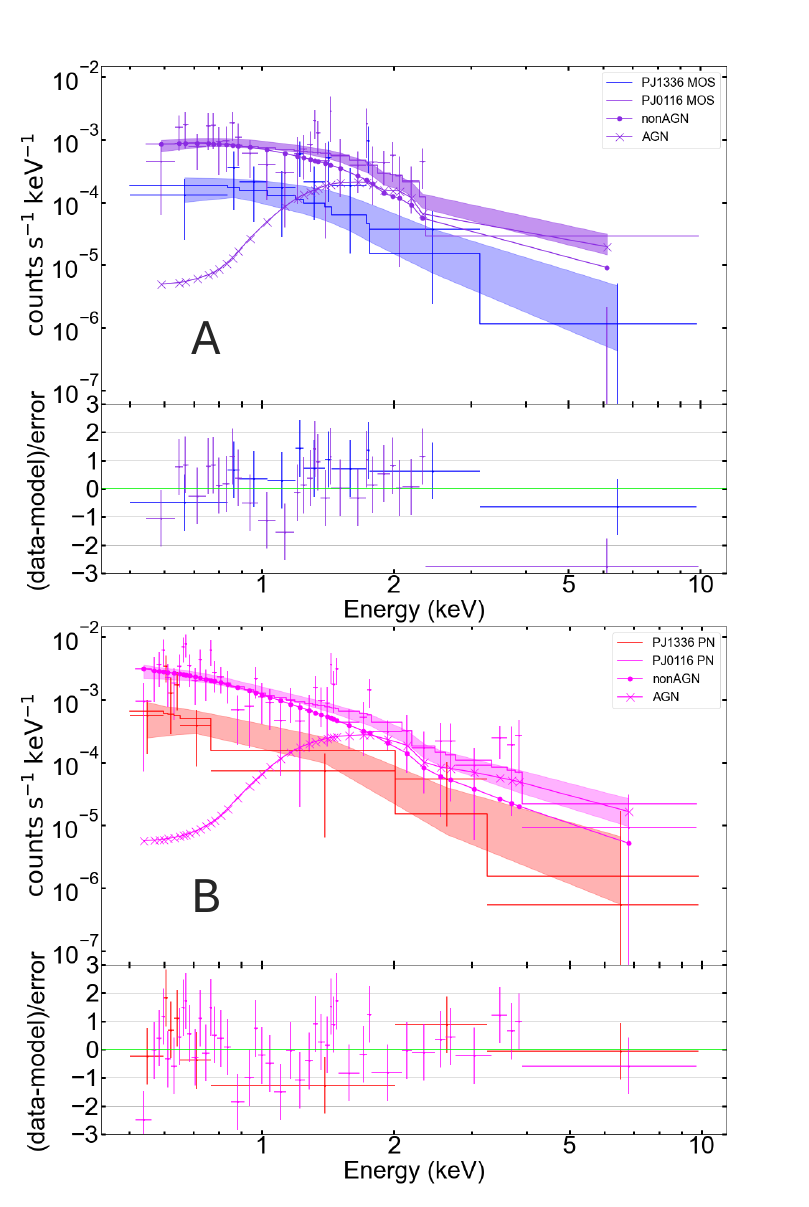}
}
\caption{
Joint fit to the \xsb\ and \xsa\ MOS spectra (A) and pn spectra (B) (Table~\ref{t:spec}). The X-ray spectral model is a power law for \xsa, representing the collective contribution of the HMXBs in the galaxy, while for \xsb\ a separate power law is included to account for the contribution from the AGN contribution. These two components are also separately plotted. 
}
\label{f:f15}
\end{figure}

\begin{figure*}
\centerline{
\includegraphics[width=1.0\linewidth,angle=0]{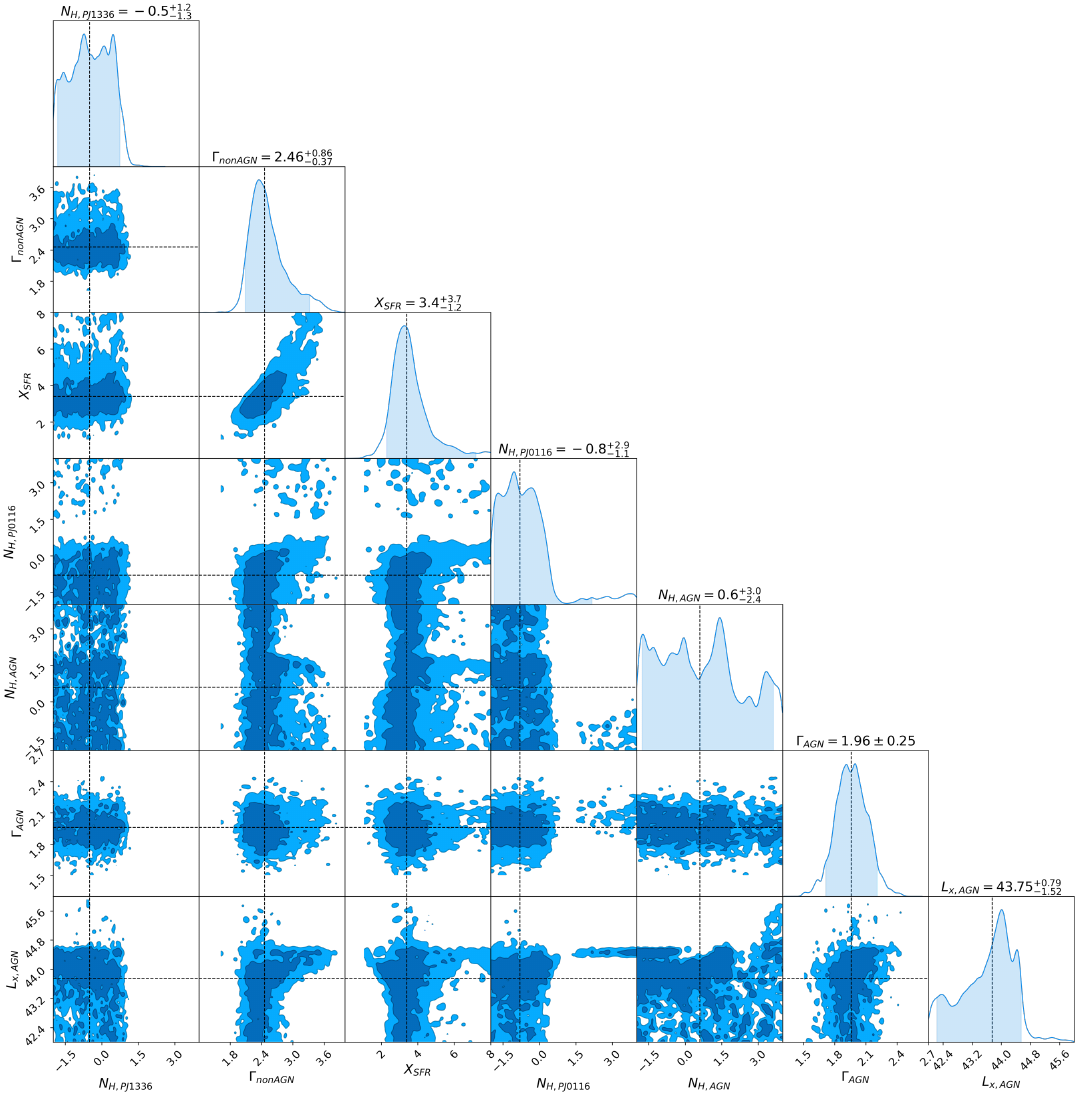}
}
\caption{
Corner plot of the best-fit model parameters from the joint-fit to the spectra of \xsb\ and \xsa\ (Table~\ref{t:spec}). 
}
\label{f:f16}
\end{figure*}

We further jointly fit the non-AGN spectral components of the \xsb\ and \xsa\ to better constrain \xp\ and isolate the AGN contribution in \xsb. The reason for doing this two-source joint fit (instead of a joint fit to the three sources) is that the AGN contribution in \xsb\ is relatively small, while the \xsc\ spectrum is more AGN-dominated.
Furthermore, the lopsided highest X-ray intensity contour seen in Fig.~\ref{f:f10}A is consistent with the AGN being at the nucleus of \xsb, forming the two lensed peaks 1a and 1b in Fig.~\ref{f:f4}A \citep{Kamieneski2023}. 
 The results of the joint fit are included in Table~\ref{t:spec}, as well as shown in Figs.~\ref{f:f15} and \ref{f:f16}. The fitted \xp\ is consistent with that from the fit to the \xsa\ spectrum alone but the uncertainty interval is significantly tightened. The obtained \xp\ is 3.4 [with the 90\% confidence uncertain interval of (2.2,7.1)]. 

The presence of an AGN in the \xsc\ field is now evident. Its X-ray emission in projection extends much further away from the lensed DSFG towards the southwest, where an apparent mid-IR counterpart of this AGN is found (Appendix~\ref{a:PJ105322-IRAC}).  We decompose the \xmm\ spectra of \xsc\ using two power law models for the AGN and HMXB contributions and assuming that the redshift of the AGN is the same as that of the DSFG (Fig.~\ref{f:f11}C and \ref{f:f14}; Table~\ref{t:spec}). The fitted \xp\ of 14.6 (8.7,23.8) is surprisingly high, although the reliability of this measurement cannot be fully judged with the limited quality of the \xmm\ spectral data. The AGN redshift assumption should have little direct effect on the power-law index of the AGN component, but would change the best-fit values of its intrinsic absorption and luminosity if its actual redshift is different, which in turn could affect the constraint on \xp. Future anticipated Gemini-N observations will help to measure the redshift of the AGN directly, and further spatial/spectral analysis of \xsc\ will be reserved for \citet{Diaz2023}.  What we can conclude here is that the spectral data of \sous\ are all consistent with an enhanced \xp. 

\section{Discussions}\label{s:dis}

The goal here is to better understand our results and their implications in a broader context. We first compare our \xp\ estimate with existing ones based primarily on local star-forming galaxies and then examine the origin of the enhanced \xp\ in the \sous. We also discuss the implications of the AGN detections in \xsb\ and  \xsc\ for the AGN and extreme SF co-evolution.

\subsection{Uncertainties in the \xp\ Estimate}
\label{ss:dis-comp-multi}

Our best constraint on the non-AGN component comes from the joint fit of the \xsb\ and \xsa\ \xmm\ spectra (Figs.~\ref{f:f15} and \ref{f:f16}; Table~\ref{t:spec}). The power law index of the non-AGN component, $\Gamma_{HMXB} = 2.46$ (2.09, 3.32), is consistent with what is expected from the collective X-ray spectrum of HMXBs, which can be characterized by a power law with $\Gamma_{HMXB} = 2.1\pm 0.1$ in the 0.25-8~keV range, after accounting for their intrinsic diversity of emission and absorption properties \citep{Sazonov2017}.  The overall contributions from hard, soft, and supersoft sources are comparable in the energy range. While the soft part (below 2 keV) of the spectrum is mostly due to soft and supersoft sources, classical harder ULXs dominate at energies above a few keV, which is most relevant here since we are observing the rest-frame $\gtrsim 2$~keV emission of \sous\ at $z > 2$. The spectrum is expected to steepen with increasing energy, as typically seen for individual ULXs \citep[e.g.,][]{Bachetti2013,Brightman2018,Walton2020}.

Our best estimate of the \xp\ factor (Table~\ref{t:spec}) is a factor of $\sim 3.4$ larger than expected from $X_{\sc HMXB,r}  $ (Fig.~\ref{f:f1}). 
A possible systematic uncertainty is in the X-ray absorption correction. The \xp\ estimate is correlated with the absorption in the X-ray spectral fits. 
Our best-fit intrinsic $N_{H,HMXB}$ (Table~\ref{t:spec}) is quite moderate; the absorption correction for \xp\ or $\mu L_X$ is only 3.1\% and 0.8\% for \xsb\ and \xsa. Our X-ray spectral modeling approximates the absorption as being the foreground of the X-ray emission. Any non-uniformity of the absorption towards the star-forming regions in a galaxy should lead to an underestimate of the actual $N_{H,HMXB}$ and hence \xp\ in our spectral modeling. Accounting for this underestimate, which is typically hard to quantify, would make \xp\ even higher. For \xsb, we can compare our X-ray absorption estimate with the optical reddening measurements via the VLT/ERIS spectroscopy: $E(B-V)_* = 0.49 \pm 0.10$ and $E(B-V)_{neb} = 0.99 \pm 0.21$ towards stars and HII nebulae, respectively \citep{Liu2023}. These measurements, together with the conversion of $N_H =4.8 \times 10^{21} E(B-V){\rm~cm^{-2}~mag^{-1}}$ for the ISM in our Galaxy \citep{Bohlin1978}, give the corresponding $N_H = (2.3 \pm 0.5) \times 10^{21} {\rm~cm^{-2}}$ and $=(4.5 \pm 1.0) \times 10^{21} {\rm~cm^{-2}}$, which are slightly larger than $N_{H,HMXB} \sim 1.6 \times 10^{21} {\rm~cm^{-2}}$ obtained from our X-ray spectral modeling (Table~\ref{t:spec}). Such uncertainty at the $N_{H,HMXB} \sim 10^{21} {\rm~cm^{-2}}$ level has a negligible effect on our measurement of \xp\ (especially in the rest-frame energy range above 1.5 keV), compared to other systematical and statistical uncertainties in our spectral analysis. Therefore, we conclude that the large \xp\ from our modeling of \xsb\ and \xsa\ cannot be due to an overestimation of the X-ray absorption.

In fact, we could have underestimated the absorption, which may be caused by the presence of soft X-ray emission in addition to the power-law components used in our spectral modeling. Such a soft excess can arise in an AGN. The origin of the soft excess is one of the major open questions in AGN research \citep[e.g.,][]{Turner2009,Nandi2023}.  Common in Type 1 AGNs \citep[e.g.,][]{Piconcelli2005}, the soft excess can be comparable to the luminosity of the AGN power-law component in the rest-frame 0.5-2 keV band and can typically be fitted with a thermal model with a characteristic temperature $\sim 0.1-0.3$~keV. Its expected peak emission is thus clearly outside the rest-frame energy range used in our spectral analysis. However, the presence of the high-energy tail of the excess could potentially lead to an underestimation of the X-ray absorption. 

A similar effect can be expected from diffuse soft X-ray emission around an AGN and in the general ISM.  Soft X-ray emission from the general ISM can be expected in star-forming galaxies, primarily because of stellar feedback from massive stars via fast stellar winds and core-collapsing SNe. These diffuse soft X-ray emission components are again typically important at photon energies $< 1.5$~keV, which are not included in our spectral analysis.  Similarly, the interaction between the energetic feedback of an AGN (via jets, winds, and radiation) and its surrounding medium can produce an extended emission that tends to be soft. The physical scale of the emission is typically significant on sub-kpc scales \citep[e.g.,][]{Travascio2021,Fabbiano2022}.  

Extended hard X-ray emission can also be produced by AGNs, mostly due to scattering of AGN radiation (including both reflection and fluorescence of photons) by the ISM, although heating by fast shocks can occasionally be considerable.  Such emission typically accounts for $\sim 1\%$ of the intrinsic X-ray luminosities of local AGNs \citep[e.g.,][]{Ricci2017b}. For such an emission to explain the enhanced \xp, the AGNs in \xsa, \xsb, and \xsc\ need to have apparent luminosities $\sim 3, 4$, and 10 $\times 10^{46} {\rm~erg~s^{-1}}$, much greater than those inferred from our spectral analysis (Table~\ref{t:spec}).  However, the emission can be important for AGNs that are strongly obscured.  Since the dense clouds responsible for the scattering are typically more concentrated at smaller galactocentric radii, the intensity of this emission is expected to decrease rapidly with off-AGN distance. Indeed, such emission has been detected up to several $\times 10^2$~pc radii around nearby Compton-thick AGNs and with total 3-7 keV luminosities up to several $\times 10^{40}{\rm~erg~s^{-1}}$ \citep{Fabbiano2022,TrindadeFalcao2023}. The angular to physical scale conversion is 7.659, 8.46, or 7.433~kpc per arcsecond at redshifts 3.25, 2.12, or 3.55 of \xsa, \xsb, or \xsc. Therefore, even taking into account the lensing amplification (e.g., 6.5 and 6.0 for the AGN in \xsb), a kpc-scale feature in the galactic core region would likely be observed on the sub-arcsecond scale in the image plane, smaller than the resolved X-ray emission extent. 

It may be possible that the bulk of the enhanced X-ray emission arises from the hot intergalactic medium (IGM) surrounding individual \sous. The situation may be similar to the Spiderweb protocluster (J1140-2629) at $z = 2.156$), for which a  \chandra\ ACIS-S observation (700 ks) reveals the presence of significant extended X-ray emission out to a radius of $\sim 12^{\prime\prime}$  (100 kpc), as well as an AGN and radio jets \citep[e.g.,][]{Tozzi2022,DiMascolo2023}. The luminosity of the thermal emission is comparable to those of the non-AGN component of  the \sous\ reported here.  However, in this scenario one should not expect any significant X-ray absorption towards the diffuse emitting regions, which seems to be marginally inconsistent with the intrinsic $N_H$ measurements (or $N_{H,HMXB}$ values in Table~\ref{t:spec}).  Furthermore, protoclusters like the Spiderweb are predicted to be very rare at $z > 2$ \citep[e.g.][]{Saro2009}. It is therefore unlikely that our target \sous, especially those \xsa\ and \xsc\ with $z > 3$, are all inside such protoclusters. Nevertheless, the protocluster scenario of enhanced X-ray emission in the \sous\ needs to be explored in future work.

Finally, the large \xp\ in \sous\ is in great contrast to the X-ray deficiency observed in nearby (U)LIRGs \citep[e.g.,][]{Persic2007,Iwasawa2009,Lehmer2010,Torres-Alba2018}. This is presumably due to the fact that the SF in \sous\ is typically more distributed than in (U)LIRGs, which minimizes both the severe absorption and youth/rapid cluster destruction effects of compact starbursts on the X-ray emission (\S~\ref{sss:intro-ratio}). In particular, both soft X-ray contamination and absorption effects should be much less important in our X-ray luminosity or \xp\ estimate made in the rest-frame energy range above 1.5~keV. As discussed above, the moderate X-ray absorption toward star-forming regions in the \sous\ seems to be reasonably well accounted for in our spectral modeling. Taking all this into account, we conclude that \xp\ is probably indeed $\sim 3$ or could be even higher in the \xsc\ case. 

\subsection{Origin of the \xp\ 
Enhancement}\label{ss:dis-Xf}

It has been shown that a significant non-AGN $L_X$ or \xp\ enhancement can be caused by the very low metallicity ($\lesssim 10\%$ solar, due mostly to low stellar mass loss) and/or the youth of a stellar population 
\citep[e.g.,][]{Fragos2013,Gilbertson2022}. However, this scenario does not apply to \sous. They tend to have extreme but distributed SF \citep{Swinbank2015,Rujopakarn2016,Iono2016,Kamieneski2023}, instead of a very localized starburst as typically seen in nearby (U)LIRGs \citep[e.g.,][]{Diaz-Santos2010}. Such distributed SF should typically have lasted longer than the dynamic time scale of a galaxy ($\sim 10^8$~yr). Although the SF could have started in a low metallicity environment, the chemical enrichment is expected to have a short time scale in \sous; so their metallicity is expected to be close to solar, or at least not particularly low, consistent with their dusty nature. Indeed, the metallicity of \xsb\ is shown spectroscopically to be slightly supersolar \citep{Liu2023}. In comparison, HMXBs, which should dominate the non-AGN component in such galaxies and trace the very recent star formation history over a period probably mostly shorter than the dynamic time scale of our target \sous. Therefore, we don't expect the large \xp\ to be due to the very low metallicity or youth of their stellar population.

An \xp\ enhancement could also be caused by an underestimation of $\mu$SFR. For each of our target \sous, $\mu$SFR (Table~\ref{t:targets}) is converted from the IR (rest-frame 8-1000 $\mu$m) luminosity derived from the best fit of the spectral energy distribution (SED) \citep{Berman2022,Liu2023}. The accuracy of the SED depends mostly on the photometry, the uncertainty of which is largely systematic and difficult to quantify \citep{Berman2022}, but could be up to 30-40\% or even more. For \xsb, \citet{Liu2023} have recently obtained an alternative estimate of $\mu$SFR $=(2.11\pm0.45)  
\times 10^4 {\rm~M_\odot~yr^{-1}}$ from the H$_\alpha$ emission corrected for extinction by the Balmer decrement H$_\alpha$/H$_\beta$ ratio. 
This estimate is only slightly less than $\mu$SFR$= (2.51\pm0.65) 
\times 10^4 {\rm~M_\odot~yr^{-1}}$ from the SED fit (Table~\ref{t:targets}) and within their quoted 1$\sigma$ error bars. An under-estimate of the $\mu$SFR is expected from the H$_\alpha$ emission, since the decrement correction most likely underestimates the extinction because of its mixing with the emission on the galaxy scale. Therefore, the comparison gives us some confidence in concluding that our SED-inferred $\mu$SFRs of the \sous\ are not systematically underestimated to explain the large \xp\ \citep[see also ][]{Mineo2012}.

One may wonder if the differential lensing of the \sous\ might play a role in the measurement of \xp. We have so far assumed that $L_X$ and $L_{IR}$ (or other SFR tracers) of a galaxy are magnified equally, with their ratio essentially unchanged in the image plane of the lensed galaxies {\S~\ref{ss:intro-lensing}). However, the assumption may not hold if the dynamical formation of HMXBs is important in \sous, which may be expected in dense clusters and probably in galactic nuclear bulges \citep[e.g.,][]{Garofali2012,Rangelov2012,Kremer2020,Rizzuto2022}. If this is the case, then the \xp\ depends on local clustering properties and is thus intrinsically non-uniform across a galaxy. This non-uniformity can, in principle, be enhanced by differential lensing magnification. X-ray emission from a dense stellar cluster, if located at the lensing caustic and much more compact than its surrounding far-IR emission region, could be disproportionally enhanced. However, because the source plane area covered by very high magnification ($\mu> 20$) is very small, we don’t expect that such a differential magnification would systematically enhance \xp. 
 
We do suspect that the dynamically formed population of HMXBs itself is responsible for the enhancement of the non-AGN X-ray emission in \sous. This scenario is similar to that used to explain the population of LMXBs observed in nearby globular-cluster-rich elliptical galaxies \citep[e.g.,][]{Gilfanov2022}. Dynamical interactions can cause mass segregation, tidal capture and disruption, and multibody exchange and binary mergers (e.g., via Lidov-Kozai cycles or by stellar collisions; \citet{Kozai1962,Eggleton2001,Baumgardt2011}) in dense stellar clusters \citep[e.g.,][]{Fabian1975,Hills1976,Stone2017} and possibly in galactic nuclear bulges \citep[e.g.,][]{Voss2007,Zhang2011}. More massive stars, more compact stellar remnants (neutron stars and BHs), and more X-ray binaries are thus expected to form \citep[e.g.,][]{Stone2017,Rizzuto2023}. In extremely dense stellar environments (e.g., when the central number density of stars $ \gtrsim 10^7 {\rm~pc^{-3}}$), initial stellar mass BHs may even grow into IMBHs via tidal capture and disruption events \citep[e.g., ][]{Rizzuto2023,Stone2017,DiMatteo2023}, which could explain BHs with masses of a few 10s (even $\gtrsim 100 {\rm~M_\odot}$), detected by the LIGO-Virgo-KAGRA interferometers \citep[e.g.,][]{Mahapatra2021}. Such formed IMBHs have also been considered to be potential seeds of SMBHs. Furthermore, the dynamical interactions can also kick out a good fraction of the binaries from the clusters, eventually forming X-ray binaries that are observed outside individual clusters. In short, the dynamical effect in young and intermediate-age dense clusters \citep[e.g.,][]{Kremer2020,Rizzuto2022,Hunt2023} is expected to produce disproportionately more compact stars (neutral stars and BHs) and more massive ones and more in accreting binaries, which can be responsible for generating more non-AGN X-ray emission and hence the large \xp. Indeed, an indication for the increasing \xp\ with the surface SFR, traced by the observed sub-mm emission, is seen in \xsc. The  \xmm\ image of this \sou\ is sufficiently resolved to allow for a {\sl spatial} decomposition of the non-AGN and AGN components (Fig.~\ref{f:f10}B). A preliminary analysis of the image shows an excess of the X-ray emission associated with the northeast CO intensity peak of the lensed ring of the DSFG, which can naturally be explained by a surface-SFR-dependent enhancement of the \xp. The quantification of this dependence will be attempted in an upcoming paper by \citet{Diaz2023}.

We expect that \sous\ are the sites where the dynamical effects manifest greatly in generating a large population of luminous HMXBs such as ULXs. It has long been proposed that gas-rich galaxies such as \sous\ tend to have higher molecular gas densities, resulting in higher SFRs and clustering efficiencies, and can form cluster populations that extend to higher initial cluster masses \citep[e.g.,][]{Adamo2020}. In particular, galaxy mergers are considered to be the most efficient producer of dense clusters. A recent high-resolution simulation of galaxy mergers \citep{Li2022} shows that the final coalescence of two large gas-rich galaxies is very turbulent and clumpy, which can lead to the formation of a large number of dense clusters whose masses dominate the total SFR. This can be seen in a PASSAGES DSFG at $z=2.24$ in the PLCK G165.7+67.0 galaxy cluster, for which spatially-resolved {\sl JWST} NIRSpec spectroscopy shows the galaxy to separate into several interacting high mass clumps \citep{Frye2023b}. There is also a strong positive correlation between the efficiency of cluster formation and the surface SFR, as well as the turbulent gas pressure. Thus it is natural to predict that the BH merger rate in stellar clusters is highest at cosmic noon \citep{DiMatteo2023}, where the population of \sous\ also peaks. Furthermore, starbursts in \sous\ seem to be similar to other clumpy star-forming galaxies observed at high-$z$ \citep[e.g.,][]{Elmegreen2009,Swinbank2015,Rujopakarn2016,Iono2016,Kamieneski2023} and substantially more spatially distributed than in local (U)LIRGs \citep{Kamieneski2023}, which minimizes the effects of the extreme extinction/X-ray absorption and stellar cluster destruction (see \S~\ref{sss:intro-ratio}). 

The stellar dynamical effect has also been proposed to explain other properties of high-$z$ galaxies. Perhaps most notable are the anomalously high [N/O] ratios observed by {\sl JWST}, similar to those known in globular clusters of the Milky Way \citep{Belokurov2023}. Stellar collisions or other dynamical processes in such clusters can naturally increase the formation of close binaries, in addition to the formation of very massive stars, which may be responsible for high UV continuum emission observed \citep[e.g.,][]{Cameron2023}. Thus, newly synthesized nitrogen can be exposed and ejected into the ISM \citep[e.g. via Wolf-Rayet phases;][]{Harrington2019}. Massive close binaries could further evolve into HMXBs. We conclude that stellar dynamics may play a very important role in determining the amount and spectrum of the radiation from stars in extreme star-forming galaxies, which are common at high-$z$.

\subsection{AGN and SF co-evolution in \sous}\label{ss:dis-AGN}

Our detection of AGN in \xsb\ and in the immediate vicinity of \xsc\ was not expected based on our selection criteria stated in \S~\ref{s:targ}. The PASSAGES galaxies were selected to be apparently starburst-dominated. This selection is based on the comparison in the mid- and far-IR color-color diagrams for starburst/AGN templates \citep{Harrington2016,Harrington2018} and the degree of radio and/or mid-IR excess in the SED fits \citep[e.g.,][]{Yun2001,Berman2022}. Two of our three \xmm\ targets, \xsb\ and \xsa, are included in the study by \citet{Kamieneski2023}. They show that both have radio/FIR correlation parameters $q_{FIR}$ consistent with the median for their selected PASSAGES DSFGs ($q_{FIR} \sim 2.4$) and apparently above the $q_{FIR} < 1.8$ threshold for an AGN-powered object (see their Table 4 and Fig. 9).
Our X-ray detection of AGN in \xsb\ and \xsc\ thus indicates that AGN may be quite common in \sous, although this is not evident in wavelength bands other than X-ray. Now that we know of the presence of the AGNs, let us examine their multi-wavelength properties, which helps to determine their nature and potential role in the AGN and SF co-evolution. 

The association of an AGN with the DSFG in \xsb\ is compelling, especially now with spatially resolved multi-wavelength spectroscopic data. Although the dust continuum and CO emissions do not peak at the center of the DSFG, the radio continuum emission does (Fig.~\ref{f:f4}A), and most likely represents this AGN. This scenario is consistent with a high [N{\sc ii}]$\lambda$6584/H$\alpha$ ratio of $1.35\pm 0.20$ at the center (compared to the galaxy-wide [N{\sc ii}]$\lambda$6584/H$\alpha$ ratio of $0.36\pm 0.04$) measured with the near-IR ERIS IFU data, as well as with the X-ray spatial and spectral properties of the galaxy (e.g., \S~\ref{s:res}).
The apparent lack of dust emission in the central region of the DSFG as seen in its source plane sub-mm image suggests that much of the gas may have been ejected from the central region, probably by the AGN \citep[e.g.,][]{Stacey2022}. The northeast and southwest images of the AGN have lensing magnifications of 6.5 and 6.0 in the \hst\ image \citep{Liu2023}. Therefore, the total lensing magnification factor is about $\sim 13$. With this factor, we infer that the intrinsic luminosity of the AGN is $\sim 10^{43} {\rm~erg~s^{-1}}$. However, the uncertainty in the luminosity is large, mainly because of the uncertainty in the foreground absorption, which could be enhanced by the presence of the AGN torus \citep[e.g.,][]{Yamada2023}. 

The presence of an AGN in the vicinity of \xsc\ was first indicated by its X-ray morphology and spectrum (\S~\ref{s:res}). A follow-up multi-wavelength examination described in Appendix~\ref{a:PJ105322-IRAC} led to the identification of this AGN based on its near- to mid-IR colors. It is located at R.A./Dec. (J2000) $=10:53:21.75/60:51:41.4$, or about $6.6^{\prime\prime}$ offset from the lensed DSFG (Fig.~\ref{f:fa1}). The AGN appears point-like in the \hst\ 1.6-mm image. To firmly establish the relationship between the AGN and the DSFG, we will first need to measure the redshift of the AGN.

If the AGN is indeed at the same redshift as \xsc, then we will have to seriously consider several intriguing questions. One of them is whether the AGN was ejected from the DSFG \citep[e.g.,][]{vanDokkum2023,Smole2023}. Such an ejection can occur during the merger of a binary SMBH. Gravitational radiation is emitted anisotropically due to asymmetries in the merger configuration. 
This anisotropic radiation leads to a gravitational wave kick, or recoil velocity, as large as a few $10^3 {\rm~km~s^{-1}}$ \citep[e.g.,][]{Campanelli2007,Smole2023, DiMatteo2023}. At such a speed, the separation between the AGN and the DSFG can probably be achieved well within $10^8$~yr. This ejection scenario is consistent with the apparent lack of evidence for a counterpart to the AGN in the existing sub-mm continuum and CO data, indicating the dearth of the ISM. 
Alternatively, the AGN and DSFG may represent a pair of two rare galaxies, separated by $\lesssim 50$~kpc in projection, in the same forming galaxy cluster. Interestingly, the DSFG has a similar color, most likely due to the strong reddening of stellar light. 

 While a detailed analysis of \xsc\ will be presented in a later paper \citep{Diaz2023}, we note that such associations of strongly lensed extreme DSFGs and AGNs provide us with excellent laboratories for understanding similar systems that are mostly unlensed and thus difficult to study in a spatially resolved fashion. One example is the unlensed HATLAS J084933.4+021443 \citep[][see also the discussion in \S~\ref{sss:intro-AGN}]{Ivison2019}. Existing studies also show that extreme SFRs ($\gtrsim 300 {\rm~M_\odot~yr^{-1}}$) tend to be seen in the hosts of AGNs with intermediate X-ray luminosities ($10^{42.5} {\rm~erg~s^{-1}} \lesssim L(8-28 {\rm~keV}) \lesssim 10^{44} {\rm~erg~s^{-1}}$), and that many dusty and fainter AGNs remain to be detected in distant galaxies ($z \gtrsim 2$; e.g., \citet{Barger2019}). Furthermore, \citet{Kirkpatrick2017} point out that there is a rare population of weakly obscured AGNs that appear to coexist with galaxies having large amounts of cold gas or extreme SFR, which may represent a short-lived phase in the co-evolution of SMBHs and galaxies (see also \cite{Sokol2023}). The strongly lensed \sous\ allow for the spatially resolved study of such extreme DSFGs associated with AGNs, especially with high angular resolution X-ray observations which could be provided by {\sl Chandra} and potentially {\sl AXIS} (Advanced X-ray Imaging Satellite -- a NASA concept probe mission) \citep{Mushotzky2019}.

\section{Summary, Conclusions and Future Prospects}\label{s:sum}

We have presented the \xmm\ observations of three \sous\ (Table~\ref{t:targets}) in the redshift range $z=2.12-3.55$, selected from the PASSAGES sample of strongly lensed DSFGs, the ratio to sub-mm SEDs of which show no sign for AGNs. The primary goal of our \xmm\ observations is to measure the collective X-ray emission of HMXBs and to constrain the \xp\ factor in our target \sous. The combination of the lensing magnification and the deep \xmm\ observations, together with the extreme {\sl intrinsic} infrared brightness of the \sous, has allowed us to detect the X-ray emission from each of them. This detection is complemented by the analysis and modeling of multi-wavelength data, some of which are presented for the first time.  We have obtained the following main results and conclusions:

\begin{itemize}
\item The 0.5-8~keV luminosity of \xsa\ is significantly greater than that expected from the reference calibration $X_{\sc HMXB,r}  $ of $3.9 \times 10^{39} {\rm~erg~s^{-1}/(M_{\odot}~yr^{-1}})$. Its X-ray spectrum is consistent with that of ULXs, which are expected to dominate the HMXB emission. This, together with the absence of multi-wavelength evidence for an AGN in the lensed DSFG, suggests that the observed X-ray emission is primarily non-AGN in origin. 

\item The X-ray emission from \xsb\ is partly contributed by an AGN. The X-ray morphology is consistent with this AGN being at the center of the DSFG, as concluded from a multiwavelength analysis of the \sou. The DSFG consists of a central stellar spheroid with little dusty gas and a disk of extreme SF with a compact blue clump. A joint spectral analysis of the non-AGN X-ray emission from \xsb\ and \xsa\ allows us to better constrain its properties. We find that \xp\ is about 3.4 or 2.2-7.1 (90\% confidence uncertain interval) in units of $X_{\sc HMXB,r}$. 

\item The X-ray emission of \xsc\ morphologically shows the presence of an AGN. It has a near- to mid-IR counterpart, $\sim 6.6^{\prime\prime}$ (or $\sim 50$~kpc in projection) from the lensed image centroid of the DSFG component, and is largely naked, lacking dust continuum and CO emission in existing observations. A spectral decomposition of the X-ray emission from \xsc\ suggests that the AGN has a rest-frame 0.5-8~keV luminosity of $\sim 10^{46} {\rm~erg~s^{-1}}$ and is obscured with log[$N_H$ (${\rm cm^{-2}}$)] $\sim 23.7$ if it is at the same redshift as the DSFG. In this case, the AGN could be kicked out of the DSFG in the recent past (e.g. $\lesssim 10^8$~years ago), probably during a major galaxy merger. 

\item Our results suggest that AGNs may be quite common in \sous, although a study of a large sample of them is highly desirable. The strong lensing of such galaxies, as provided by the PASSAGES sample, provides a powerful tool to examine the AGN co-evolution with extreme SF.

\item Our measured \xp\ factor provides a tight constraint on the HMXB population in the \sous. The large \xp\ ($\sim 3$; Fig.~\ref{f:f1}) can naturally be explained by an enhanced population of HMXBs (due to increased numbers and BH masses) that could dynamically form in dense clusters, some of which may eventually evolve into today's globular clusters. HMXBs formed in this way also produce BH binaries, the eventual merger of which can be detected by gravitational waves. The inclusion of the dynamical formation could significantly change the HMXB population synthesis modeling \citep[e.g.,][]{Fragos2013}, which has so far neglected such an effect.

\end{itemize}

While this \xmm\ study demonstrates the scientific potential of X-ray observations of strongly lensed \sous, more work is clearly needed. For example, to further test our postulated scenario for the enhanced \xp, we need high angular resolution X-ray data, which can be provided by deep \chandra\ observations. Such data will allow a clear separation of the AGN and non-AGN components. We can also study the spatially resolved \xp\ distribution. If the dynamical formation of HMXBs is indeed important, \xp\ should statistically increase with the surface SFR or correlate with the presence of dense clusters, which can be probed by {\sl JWST} observations. Such studies will allow us to learn about important high-energy processes, in particular the co-evolution of stellar to massive BHs with extreme SF in \sous. 
\section*{Acknowledgment}
We thank the referee, Giuseppina Fabbiano, for prompt and constructive comments, which helped to improve the paper. This work is largely supported by NASA under the grant 80NSSC22K1923, based on data obtained from \xmm.
This research is also based on observations made with the NASA/ESA Hubble Space Telescope obtained from the Space Telescope Science Institute, which is operated by the Association of Universities for Research in Astronomy, Inc., under NASA contract NAS 5—26555. These observations are associated with programs GO-14653 and GO-14223. Support for program GO-14653 was provided by NASA through a grant from the Space Telescope Science Institute. Some of the data presented in this paper were obtained from the Mikulski Archive for Space Telescopes at the Space Telescope Science Institute. This paper further makes use of the following ALMA data: ADS/JAO.ALMA \# 2017.1.01214.S, as well as data from VLA and SMA. 
ALMA is a partnership of ESO (representing its member states), NSF (USA) and NINS (Japan), together with NRC (Canada), MOST and ASIAA (Taiwan), and KASI (Republic of Korea), in cooperation with the Republic of Chile. The Joint ALMA Observatory is operated by ESO, AUI/NRAO and NAOJ. The National Radio Astronomy Observatory is a facility of the National Science Foundation operated under cooperative agreement by Associated Universities, Inc.
The Submillimeter Array is a joint project between the Smithsonian Astrophysical Observatory and the Academia Sinica Institute of Astronomy and Astrophysics and is funded by the Smithsonian Institution and the Academia Sinica. E.F.-J.A. acknowledges support from UNAM-PAPIIT project IA102023, and from CONAHCyT Ciencia de Frontera (project ID: CF-2023-I-506). 

\section*{DATA AVAILABILITY}

The X-ray data presented here are available in the \xmm\ data archive (https://www.cosmos.esa.int/web/xmm-newton/xsa). Other processed data products underlying this article will be shared on reasonable request to the authors.
\bibliographystyle{mnras}
\bibliography{ms-V2.bbl}

\appendix 
\section{Submillimeter Array Observations (SMA) of \xsa\ and \xsc:}\label{a:sma}
We present data from two separate SMA observations for \xsa\ and \xsc\ to complete the mm-wavelength perspective of these objects in the context of the \xmm\ observations. \xsa\ was observed in the extended configuration 2022 June 12 (6 antennas; PWV$<$2.5~mm) during the 2021B semester (Project ID: 2021B-S011; P.I. Harrington, K.) using the SWARM correlator and dual receiver tuning at $\sim$243 GHz and 343 GHz. The data were calibrated using the latest SMA pipeline\footnote{\url{https://github.com/Smithsonian/sma-data-reduction.git}} implementation in CASA and will be further detailed in a forthcoming manuscript \citep{Harrington2023} analyzing the CO(9-8) and [N{\sc ii}]205~$\mu$m emission lines. \xsc\ was observed in the extended configuration 2016-Nov 22 (8 antennas; PWV$<$3.5\,mm) during the 2016B semester (Project ID: 2016B-S062; P.I. Yun, M.) using the SWARM correlator during its initial upgrade to wider bandwidth capabilities, here tuned to $\sim$200 GHz and 220\,GHz. The data were calibrated using the \textit{MIR} package and further converted to a CASA measurement set before imaging. More information will be presented in a detailed multi-wavelength analysis for \xsc\ including this SMA data set (covering CO(8-7) and CO(9-8) emission lines) \citep{Diaz2023}.

\begin{figure*}
\centerline{
\includegraphics[width=1\linewidth,angle=0]{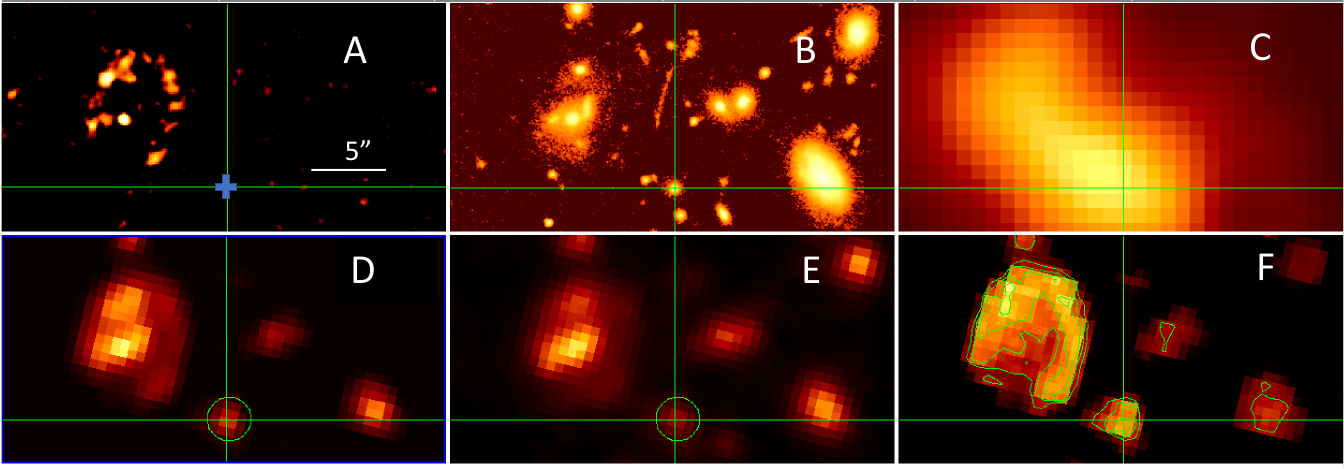}}
\caption{
Multi-wavelength view of the \xsc\ field:
{\sl VLA} 6 GHz (A), \hst\ WFC3/IR F160W (B), and \xmm\ 0.3-7~keV (C). {\sl Spitzer}/IRAC 4.5~$\mu$m (D), 3.6~$\mu$m (E), and the 4.5~$\mu$m to 3.6~$\mu$m intensity ratio (F), respectively. The ratio contours in panel F are at 0.7, 1, 1.3, and 1.6. 
}
\label{f:fa1}
\end{figure*}

\begin{figure}
\centerline{
\includegraphics[width=1\linewidth,angle=0]{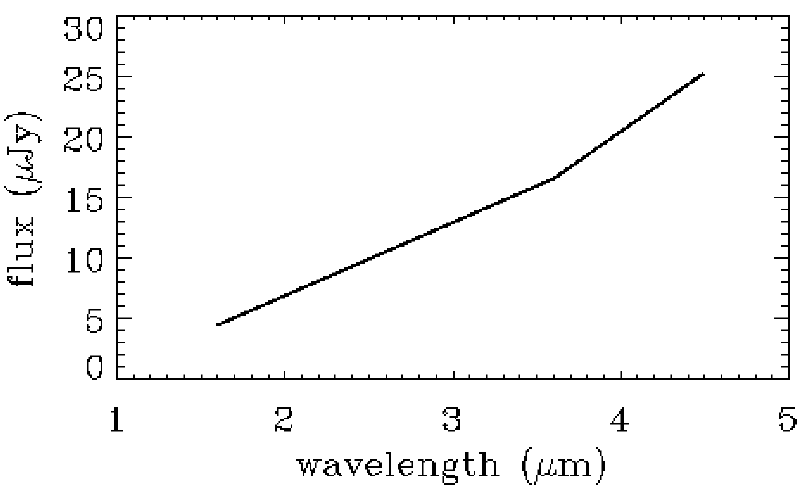}}
\caption{
NIR-MIR spectral shape of the AGN candidate in the vicinity of \xsc. This shape is estimated from the flux densities observed in the \hst\ WFC3/IR F160W and the {\sl Spitzer}/IRAC 4.5~$\mu$m, 3.6~$\mu$m bands. 
}
\label{f:fa2}
\end{figure}

\section{\Spitzer\ / IRAC data analysis of \xsc}\label{a:PJ105322-IRAC}
In the preparation of the observing program for \xsc, we find that \Spitzer/IRAC data of this target\footnote{https://sha.ipac.caltech.edu/applications/Spitzer/SHA/} are useful and are not published. We here present a preliminary analysis of the data in the multi-wavelength context of this \sou.

Our primary interest is in the mid-IR properties of the AGN candidate source southwest of the DSFG, as indicated in the \xmm\ and \hst\ images (Figs.~\ref{f:f10}B and \ref{f:fa1}B). We clearly see the counterparts of the source in the IRAC 4.5~$\mu$m and 3.6~$\mu$m images (Fig.~\ref{f:fa1}D-F). We estimate its specific energy flux to be $14.8\pm0.2$ and $23\pm2 {\rm~\mu Jy}$ in the IRAC 3.6 and 4.5 \micron\ bands. The color [3.6]-[4.5] (AB)$ = 0.48\pm0.09$, corresponding to a power-law SED $f_\nu \propto \nu^{-2}$, is consistent with the AGN nature \citep[e.g., Fig.~\ref{f:fa2};][]{Stern2005,Assef2018}. 

The IRAC data also show an apparent counterpart to the DSFG with a red color (similar to that of the AGN), indicating strong dust reddening of the underlying stellar light. The DSFG and AGN may represent two members of a proto-cluster of galaxies, which may further include a linear feature off from both DSFG and AGN, as seen in the \hst\ 1.1-\micron\ image (Fig.~\ref{f:fa1}B). Because no radio or dust continuum counterpart has been found for this feature, it likely represents an evolved galaxy strongly lensed by a foreground galaxy group complex (Appendix~\ref{a:PJ1053-lens-model}). 

\section{Lens Modeling for \xsc}\label{a:PJ1053-lens-model}

\begin{figure*}
\centerline{
\includegraphics[width=1.0\linewidth,angle=0]{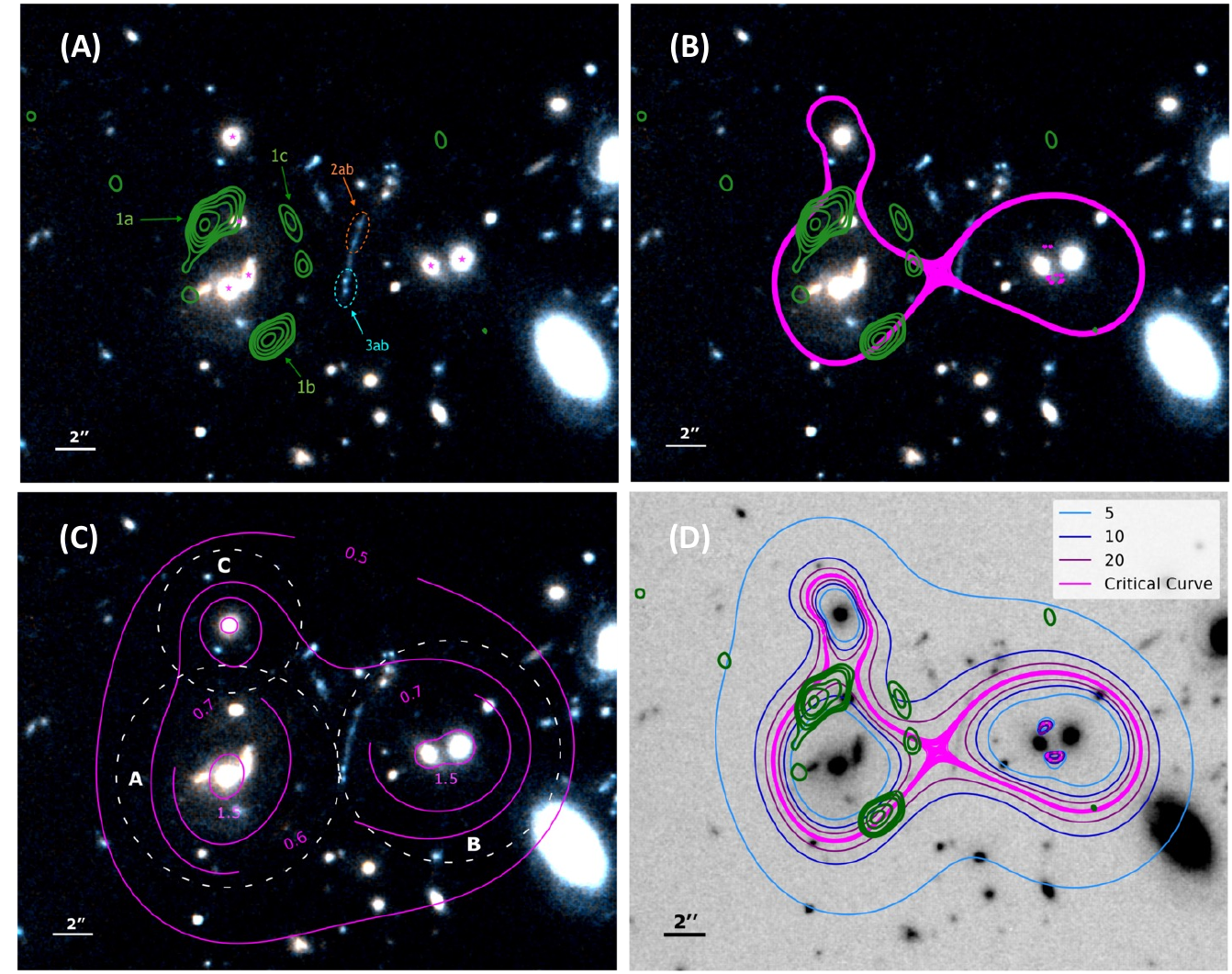}
}
\caption{
(A) The \hst\ f140W/F160W color image of the \xsc\ field, together with the SMA 850 \micron\ intensity (pink) contours. The lensed image systems 2 and 3 are labeled, while the cluster member galaxies used in the lens modeling are marked with magenta stars. (B) The same color image as in panel A, but together with the (magenta) critical curve of the lensing.
(C) The same color image as in (A), but overlaid with the marked surface mass density contours in units of the critical mass surface density. The white dashed circles enclose the regions used to estimate the gravitational masses of A, B, and C peaks as (7.7, 7.8, and 2.8) $\times 10^{12} {\rm~M_\odot~yr^{-1}}$. (D) The WFC3/IR F160W image, together with the same SMA 850 \micron\ intensity contours as in (A), but further overlaid with our best-fit magnification contours at the marked levels.
}
\label{f:fc}
\end{figure*}
The main purpose of this modeling is to provide a reasonable estimate of the lensing mass of \xsc. Fig.~\ref{f:fc} shows a complex foreground lensing system of galaxies together with the identified multiple-imaged systems. The green contour displays the SMA 850~\micron\ emission at $z=3.549$ which is lensed into a partial Einstein ring. The centroids of three emission peaks, as marked in Fig.~\ref{f:fc}A, are used for the lens modeling. We also find a long arc consisting of two sets of image systems: 2ab and 3ab (Fig.~\ref{f:fc}A), which appear to be caustic crossings and thus help constrain the lens model (Fig.~\ref{f:fc}B). However, these systems have no measurements of redshifts, which are left as free parameters in our lens modeling. These sets of lensed image systems are used to constrain the lens model.

Also marked in Fig.~\ref{f:fc} are the foreground lensing galaxies that are included in our lens-mass modeling. They are identified by searching for the red cluster sequence in a color-magnitude diagram constructed with \hst\ f140W and F160W \citep{Gladders2000}. Unfortunately, the \hst\ images are heavily contaminated by a foreground star, potentially affecting photometric and color measurements of galaxies in the field. We therefore make the foreground membership particularly strict, setting a tight color cut and including only those galaxies with the most influential lensing potential. 

We characterize the lensing mass model of the \xsc\ foreground galaxy group, following the semi-parametric Light Traces Mass (LTM) approach. The details regarding the general implementation of the LTM model, as well as the initial lens modeling of \xsc, are presented in  \citet{Frye2019}. The modeling assumes that the gravitational halo mass of the lens is traced by the observed foreground galaxies \citep{Zitrin2009,Zitrin2015}. The surface mass density profile of each galaxy is characterized by a 2-D power law with a given index or a free parameter left for the model to optimize. The 2-D mass distributions of the lensing galaxies are combined and smoothed with a Gaussian kernel of a given size, which is another fitting parameter. 
We also implement an external shear to allow some flexibility in modeling the overall ellipticity of the lens mass distribution, which introduces two additional fitting parameters, a shear amplitude and a position angle. More freedom is given to the brightest group galaxy by fitting its core radius, ellipticity, and position angle. The redshifts of the identified galaxies are also left as free parameters in the absence of reliable measurements. 

The lens model is constrained via MCMC minimization by the positions of the identified multi-lensed images, which provide strong lensing information such as parity. We find that the best-fit model reproduces the image positions of the lensed sources well with an RMS$\sim$0.3$^{\prime\prime}$ with a $\chi^2/$d.o.f $=24.02/12$. The redshifts of image systems 2 and 3 cannot be constrained well, but are consistent with that of the DSFG. Fig.~\ref{f:fc}C shows the best-fit mass distribution of the lens, which is consistent with a galaxy group morphology with distinct multiple mass peaks. The masses enclosed by the circles around the mass peaks A, B, and C are (7.7, 7.8, and 2.8) $\times 10^{12} {\rm~M_\odot}$, respectively, while the total mass enclosed by the critical curve in Fig.~\ref{f:fc}B is $\sim 1.4 \times 10^{13} {\rm~M_\odot}$. The mass of peak A is most relevant here and is listed in Table~\ref{t:targets} for \xsc. The mean magnification of the DSFG as detected by the SMA 850~\micron\ emission is $\mu\sim24$ which is also listed in the table. In addition, we present a magnification gradient map in Fig.~\ref{f:fc}D to show the differential magnification ranging from $\mu \sim $ 3-100 over the range of its images. 
The foreground lens (or eastern peak of the mass distribution) of \xsc\ has an effective Einstein radius of 5.9$^{\prime\prime}$, enclosing a mass of $\sim 1.4 \times 10^{13} {\rm~M_\odot}$ (included in Table~\ref{t:targets}). 
With the limited constraints from the three systems providing only local constraints, the global foreground structure remains poorly understood. Nevertheless, the modeling should be sufficient for our lens mass estimate (hence its X-ray contribution) and a simple reconstruction of the source plane image in the SMA band of \xsc\ (Fig.~\ref{f:f5} middle panel), as described in the main text (\S~\ref{s:data}).

\section{Estimation of the X-ray contributions from the lensing foregrounds}\label{a:lens-X-ray}
We here characterize the X-ray properties of each lens, primarily its luminosity and spectral shape, to determine the potential count rate contribution to the observed X-ray emission of the corresponding target. 
The contribution can be divided into two parts: 1) diffuse hot gas (the circumgalactic and/or intragroup medium) and 2) discrete sources, dominated by low-mass X-ray binaries (LMXBs).
Our estimation of the gas contribution is based mainly on the scaling relations given by \citet{Gaspari2019}, which are obtained from a comprehensive Bayesian correlation analysis of various galaxy/halo parameters and X-ray plasma properties over galactic to cluster scales. Relevant here are the two relations between the total mass ($M_{tot,c}$), 0.3-7~keV luminosity ($L_{X,c}$), and temperature ($T_{X,c}$) of the gas within a galactic halo core ($R_{X,c} \approx 0.15 R_{500}$), which dominates the diffuse X-ray emission of a group:
\begin{equation}
 {\rm log}\Big[\frac{M_{tot,c}(M_\odot)}{L_{X.c}(L_\odot)}\Big]=2.69(\pm0.22) {\rm log}[T_{X,c}({\rm keV})]+4.85(\pm0.08)
 \label{e:ML-T}
\end{equation}
and
\begin{equation}
 {\rm log}[L_{X.c}(10^{44}{\rm~erg~s^{-1}})]=4.53(\pm0.23) {\rm log}[T_{X,c}({\rm keV})]+4.53(\pm0.08).
 \label{e:L-T}
\end{equation}
From these two relations, we drive the following two expressions:
\begin{equation}
 L_{X.c}= (3.9 \times 10^{39}{\rm~erg~s^{-1}})\Big[\frac{M_{tot,c}}{10^{12} M_\odot}\Big]^{2.55},
 \label{e:L-M}
\end{equation}
and
\begin{equation}
 T_{X.c}(keV)= \Big[\frac{L_{X,c}}{3.0 \times 10^{41} {\rm~erg~s^{-1}}}\Big]^{0.23}.
 \label{e:T-L}
\end{equation}
 Our estimated $L_{X,c}/T_{X,c}$ values are $\sim$1.2/0.47, 0.0042/0.13, and 71/1.2 
 (in units of $10^{40} {\rm~erg~s^{-1}/keV}$) for the lensing foregrounds of \xsb, \xsa, and \xsc, respectively. 
We assumed that the gas contribution is primarily due to an optically thin thermal plasma, characterized by the \texttt{Xspec} model \texttt{clumin(apec)}, where the convolution model \texttt{clumin} conveniently calculates the intrinsic rest-frame luminosity of the contribution and replaces the normalization of the additive model (here \texttt{apec}). The intrinsic absorption inside the lens is neglected. 
Based on the above $L_{X,c}/T_{X,c}$ values, we estimate the contributions of the lensing foregrounds are 0.14\%, 0.004\%, and 1.4\% of the observed 0.5-2~keV pn count rates of \xsb, \xsa, and \xsc, respectively; the contributions are less in the MOS and/or in higher energy bands.
 
We next estimate the contribution from LMXBs of the lensing galaxies. Their average X-ray spectra are expected to be much harder than the spectra of the plasma and similar to those of the observed spectra of the \sous. We find that the integrated X-ray luminosities of the lensing foregrounds, estimated from the scaling relation $(L_{\rm X,LMXB}/10^{40} {\rm~erg~s^{-1}}) =(L_K/10^{11} L_{\rm K})$ \citep{Kim2013}, are even smaller than those from the plasma. 
For simplicity, we also neglect the LMXB contributions in our spectral modeling of the \sous.

\end{document}